\documentclass[aip,jcp,superscriptaddress,amsmath,amssymb,twocolumn,reprint]{revtex4-1}

\usepackage[version=3]{mhchem} 
\usepackage{graphicx}
\usepackage{multirow}
\usepackage{amsmath}
\usepackage{amssymb}
\usepackage{nicefrac}
\usepackage{booktabs}
\usepackage{mathtools}
\usepackage{mwe}
\usepackage{array}
\newcolumntype{m}{>{$} c <{$}}
\usepackage{color}

\def\rv{{\bf r}}
\def\tv{{\bf t}}

\def\sv{{\bf s}}
\def\kv{{\bf k}}
\def\xv{{\bf x}}
\def\beq{\begin{equation}}
\def\eeq{\end{equation}}

\def\dens{\rho}
\def\erf{\mathrm{erf}}
\def\erfc{\mathrm{erfc}}
\def\loc{\mathcal{L}}

\begin{document}     

\author{Timothy J. Daas}
\affiliation
{Department of Chemistry \& Pharmaceutical Sciences and Amsterdam Institute of Molecular and Life Sciences (AIMMS), Faculty of Science, Vrije Universiteit, De Boelelaan 1083, 1081HV Amsterdam, The Netherlands}
\author{Juri Grossi}
\affiliation
{Department of Chemistry \& Pharmaceutical Sciences and Amsterdam Institute of Molecular and Life Sciences (AIMMS), Faculty of Science, Vrije Universiteit, De Boelelaan 1083, 1081HV Amsterdam, The Netherlands}
\author{Stefan Vuckovic}
\affiliation
{Department of Chemistry, University of California, Irvine, CA 92697, USA}
\author{Ziad H. Musslimani}
\affiliation{Department of Chemistry \& Pharmaceutical Sciences and Amsterdam Institute of Molecular and Life Sciences (AIMMS), Faculty of Science, Vrije Universiteit, De Boelelaan 1083, 1081HV Amsterdam, The Netherlands}
\author{Derk P. Kooi}
\affiliation
{Department of Chemistry \& Pharmaceutical Sciences and Amsterdam Institute of Molecular and Life Sciences (AIMMS), Faculty of Science, Vrije Universiteit, De Boelelaan 1083, 1081HV Amsterdam, The Netherlands}
\author{Michael Seidl}
\affiliation
{Department of Chemistry \& Pharmaceutical Sciences and Amsterdam Institute of Molecular and Life Sciences (AIMMS), Faculty of Science, Vrije Universiteit, De Boelelaan 1083, 1081HV Amsterdam, The Netherlands}
\author{Klaas J. H. Giesbertz}
\affiliation
{Department of Chemistry \& Pharmaceutical Sciences and Amsterdam Institute of Molecular and Life Sciences (AIMMS), Faculty of Science, Vrije Universiteit, De Boelelaan 1083, 1081HV Amsterdam, The Netherlands}
\author{Paola Gori-Giorgi}
\affiliation
{Department of Chemistry \& Pharmaceutical Sciences and Amsterdam Institute of Molecular and Life Sciences (AIMMS), Faculty of Science, Vrije Universiteit, De Boelelaan 1083, 1081HV Amsterdam, The Netherlands}

\title{Large coupling-strength expansion of the M{\o}ller-Plesset adiabatic connection:
From paradigmatic cases to variational expressions for the leading terms}
\begin{abstract}
We study in detail the first three leading terms of the large coupling-strength limit of the adiabatic connection that has as weak-interaction expansion the M{\o}ller-Plesset perturbation theory. We first focus on the H atom, both in the spin-polarized and the spin-unpolarized case, 
reporting numerical and analytical results. In particular, we derive an asymptotic equation that turns out to have simple analytical solutions for certain channels.  The asymptotic H atom solution for the spin-unpolarized case is then shown to be variationally optimal for the many-electron spin-restricted closed-shell case, providing expressions for the large coupling-strength density functionals up to the third leading order. We also analyze the H$_2$ molecule and the uniform electron gas.
\end{abstract}

\maketitle

\section{Introduction}
Mixing Density Functional Theory (DFT) and Hartree-Fock (HF) ingredients is an approximation strategy that has a long history in chemistry, already starting with hybrids\cite{Bec-JCP-93a,Bec-JCP-93,PerErnBur-JCP-96,HeyScuErn-JCP-03,ZhaTru-ACR-08,JarScuErn-JCP-03,ArbKau-CPL-07} and double hybrids,\cite{Gri-JCP-06,LarGri-JCTC-10,ShaTouSav-JCP-11,SuXu-JCP-14} but also by simply inserting the HF density into a given approximate exchange-correlation (XC) density functional.\cite{GillJohPopFri-IJQC-92,OliBar-JCP-94,KimSimBur-JCP-11,KimSimBur-JCP-14,KimSimBur-PRL-13,SimSonBur-JPCL-18,VucSonKozSimBur-JCTC-2019} In these strategies, the underlying idea is to use HF ingredients to approximate the XC energy of Kohn-Sham DFT.

The reverse strategy, namely to use density functionals to model the HF correlation energy $E_c^{\rm HF}$ (also called the traditional quantum chemistry or wave function-theory correlation energy) is also a formally valid alternative.
The HF correlation energy $E_c^{\rm HF}$ has been proven\cite{HarPra-JCP-85,Lev-Erd-87,Dav-PRA-90} a long time ago to be a unique functional of the HF density, $E_c^{\rm HF}[\rho^{\rm HF}]$, and various semiempirical approximations for it were already proposed and tested before these proofs (see, e.g., Refs.~\onlinecite{LieCle-JCP-74a,LieCle-JCP-74b}). It has also been found that the Wilson-Levy functional\cite{WilLev-PRB-90} provides a decent generalised gradient approximation (GGA) of $E_c^{\rm HF}[\rho^{\rm HF}]$ for ionization energies\cite{FueSav-CPL-94} and for non-covalent interaction energies.\cite{Wal-PCCP-05,CivEtAl-CPL-07}

More recently, it has been observed that rather accurate interaction energies,\cite{FabGorSeiDel-JCTC-16,GiaGorDelFab-JCP-18} again especially for non-covalent complexes,\cite{VucGorDelFab-JPCL-18} can be obtained by modeling the  HF correlation energy $E_c^{\rm HF}$ with an interpolation between the second-order M{\o}ller-Plesset perturbation theory (MP2) and a large coupling-strength limit, which is approximated with the strong-interaction DFT functionals\cite{SeiPerKur-PRA-00,SeiGorSav-PRA-07,GorVigSei-JCTC-09} of the HF density. These interpolations can easily be corrected from their size-consistency error,\cite{VucGorDelFab-JPCL-18} and have been shown to also provide a diagnostic indicator for the accuracy of MP2 for non-covalent interactions.\cite{VucFabGorBur-JCTC-20} 
Notice that the Wilson-Levy\cite{WilLev-PRB-90} functional was also constructed by generalising Wigner's original idea\cite{Wig-TFS-38} of interpolating between weak- and strong-interaction. Thus, there seems to be an indication that non-covalent interactions can be modeled in an accurate way by using the interpolation idea to build $E_c^{\rm HF}$.

To investigate the theoretical framework behind this idea, in Ref.~\onlinecite{SeiGiaVucFabGor-JCP-2018} the large coupling-strength limit of the adiabatic connection (AC) that has the M{\o}ller-Plesset (MP) series as perturbation expansion at small-coupling (denoted here as MP AC) has been studied for the first time, proving that the leading term is determined by a functional of the HF density with a clear electrostatic physical interpretation, and also establishing an inequality with respect to the leading term of the density-fixed AC of DFT.

The aim of this work is to gain more insight in the large coupling-strength limit of the MP AC, providing new pieces of information to build better approximations. As a starting point, we look at the simplest possible system, the H atom, which we consider both in its spin-polarized and spin-unpolarized (which appears locally, in the infinitely stretched H$_2$ molecule) states. This allows us to solve exactly the large coupling-strength asymptotic equation, revealing the role of the HF exchange operator in this limit. We then show that the spin-unpolarized H atom solution allows us to write a variational estimate for the HF density functionals of the next two leading terms for large coupling strength. We also analyze the H$_2$ molecule in restricted HF (RHF) to study how it tends to twice the spin-unpolarized H atom curve as the internuclear distance goes to infinity. Finally, we look at the uniform electron gas (UEG), which provides the correct limit that the large coupling-strength HF functionals should reach when the HF density is slowly varying.

\section{Theoretical background}
We consider the MP AC, defined as the adiabatic connection that has the MP series as perturbation expansion at small coupling strengths $\lambda$ (see, e.g., Refs.~\onlinecite{Per-IJQC-18,SeiGiaVucFabGor-JCP-2018}), with the following $\lambda$-dependent hamiltonian (in Hartree atomic units, used throughout this work)
\begin{equation}\label{eq:HlambdaHF}
	\hat{H}_{\lambda}^{\rm HF}=\hat{T}+\hat{V}_{\rm ext}+\hat{J}-\hat{K}+\lambda(\hat{V}_{ee}-\hat{J}+\hat{K}),
\end{equation}
with $\hat{V}_{\rm ext}$ the (nuclear) external potential, and $\hat{J}=\hat{J}[\rho^{\rm HF}]$ and $\hat{K}=\hat{K}[\{\phi_i^{\rm HF}\}]$ the $\lambda$-independent Hartree and exchange operators (restricted or unrestricted) that are explicitly defined in terms of the HF density $\rho^{\rm HF}$ and of the occupied HF orbitals $\{\phi_i^{\rm HF}\}$, obtained from an initial standard HF calculation, i.e., by minimizing the $\lambda=1$ hamiltonian over single Slater determinants only. Notice that with our definition $\hat{K}$ is positive definite. Using the Hellman-Feynman theorem on Eq.~\eqref{eq:HlambdaHF}, one obtains
\begin{equation}
	E_{c}^{\rm HF}=\int_0^1 W_{c,\lambda}^{\rm HF}\,d\lambda,
\end{equation}
where $W_{c,\lambda}^{\rm HF}$ is defined as,
\begin{equation}\label{eq:deriv2}
    W^{\rm HF}_{c,\lambda}=\left\langle\Psi_{\lambda}\left|\hat{V}_{ee}-\hat{J}+\hat{K}\right|\Psi_{\lambda}\right\rangle+U[\rho^{\rm HF}]+E_{x}[\{\phi_i^{\rm HF}\}].
\end{equation}
containing  $U[\rho^{\rm HF}]$, which is the classical Hartree energy, $E_{x}[\{\phi_i^{\rm HF}\}]$, which is the usual HF exchange energy, and the wave function $\Psi_{\lambda}$ that minimizes the expectation value of $\hat{H}_{\lambda}^{\rm HF}$ of Eq~\eqref{eq:HlambdaHF}. This way, the small $\lambda$ expansion of $W_{c,\lambda}^{\rm HF}$ returns the MP series,
\begin{equation}\label{eq:WHFMP}
    W^{\rm HF}_{c,\lambda\rightarrow 0}=\sum_{n=2}^\infty n\,E^{{\rm MP}n}_{c}\,\lambda^{n-1}.
\end{equation}

\subsection{Summary of previous results for the $\lambda\to\infty$ limit}
In Ref.~\onlinecite{SeiGiaVucFabGor-JCP-2018}, a simple variational argument has been used to show that, when $\lambda\to\infty$, $W_{c,\lambda}^{\rm HF}$ must have an expansion formally similar to the one  of the density-fixed adiabatic connection of DFT,\cite{SeiPerLev-PRA-99,GorVigSei-JCTC-09} at least for the first two terms,
\begin{equation}\label{eq:HFSeidl}
    W^{\rm HF}_{c,\lambda\rightarrow\infty}=W^{\rm HF}_{c,\infty} +\lambda^{-1/2}\, W^{\rm HF}_{\frac{1}{2}}+O\left(\lambda^{-\frac{3}{4}}\right).
\end{equation}
Notice that in the density-fixed AC DFT case, it has been shown that\cite{GorVigSei-JCTC-09} the term after $\lambda^{-1/2}$ must be at least $O(\lambda^{-5/4})$, while one of the results of this work will be to show that in the HF case there can be a non-zero term of order $\lambda^{-3/4}$.

The way Eq.~\eqref{eq:HFSeidl} has been proven\cite{SeiGiaVucFabGor-JCP-2018} was by noticing that
in the large $\lambda$ limit the term $\lambda(\hat{V}_{ee}-\hat{J}+\hat{K})$ in Eq.~\eqref{eq:HlambdaHF} becomes dominant, and the wave function $\Psi_{\lambda}$ ends up minimizing this term alone, 
\begin{equation}\label{eq:minpsi} \lim_{\lambda\rightarrow\infty}\Psi_{\lambda}=\text{arg}\min_{\Psi}\left\langle\Psi\left|\hat{V}_{ee}-\hat{J}+\hat{K}\right|\Psi\right\rangle.
\end{equation}
Moreover, since $\hat{K}$ is a positive definite operator, the best we can do is to make it vanish as $\lambda\to\infty$. This can be achieved with a very simple variational ansatz,\cite{SeiGiaVucFabGor-JCP-2018} in which the electrons are distinguishable, and each one occupies a gaussian centered in one of the positions that minimize the multiplicative operator $\hat{V}_{ee}-\hat{J}$, 
\begin{align}\label{eq:VeeminJ}
\hat{V}_{ee}-\hat{J} & =\sum_{\substack{i,j=1 \\ j>i}}^{N}\frac{1}{|\rv_i-\rv_j|}-\sum_{i=1}^N v_h(\rv_i,[\rho^{\rm HF}]) \\
v_h(\rv,[\rho]) & =\int \frac{\rho(\rv')}{|\rv-\rv'|}d\rv',
\end{align}
seen as a function of $\rv_1,\dots,\rv_N$ whose minimum is achieved in $\rv_1^{\rm min},\dots,\rv_N^{\rm min}$,
\begin{equation}
	\Psi_\lambda^{T}(\rv_1,\dots,\rv_N)=\prod_{i=1}^N G_{\alpha(\lambda)}(\rv_i-\rv_i^{\rm min}),
	\label{eq:PsiT}
\end{equation}
where $G_\alpha(\rv)=\frac{\alpha^{3/4}}{\pi^{3/4}}e^{-\frac{\alpha}{2}|\rv|^2}$, and $\alpha \sim \lambda^{1/2}$ as $\lambda\to\infty$. Since when $\lambda\to\infty$ the square of the gaussians $G_{\alpha(\lambda)}$ appearing in Eq.~\eqref{eq:PsiT} tends to Dirac $\delta$-functions centered in different positions $\rv_i^{\rm min}$, the effect of antisymmetrisation of Eq.~\eqref{eq:PsiT} will be $O(e^{-\lambda^{1/2}})$ in the computation of the expectation values, similarly to the DFT case.\cite{GroKooGieSeiCohMorGor-JCTC-17} As $\lambda\to\infty$, thus, the expectation $\langle\Psi_\lambda^{T}|\hat{V}_{ee}-\hat{J}|\Psi_\lambda^{T}\rangle$ tends to the absolute minimum of the $3N$-dimensional function $\hat{V}_{ee}-\hat{J}$ of Eq.~\eqref{eq:VeeminJ}, which will determine the value of $W^{\rm HF}_{c,\infty}$ in Eq.~\eqref{eq:HFSeidl} once we add to it $U+E_{x}$. Since $\hat{J}$ only depends on the HF density $\rho^{\rm HF}$, the value of this minimum will be a functional of the HF density only (although the HF orbitals are also implicit functionals of the HF density,\cite{HarPra-JCP-85,Lev-Erd-87,Dav-PRA-90} here the point is that they do not appear at all in this leading term). We can also write the value of $W^{\rm HF}_{c,\infty}$ as
\begin{equation}
	W^{\rm HF}_{c,\infty}= E_{el}[\rho^{\rm HF}]+E_x[\{\phi_i^{\rm HF}\}],
\end{equation}
where the density functional $E_{el}[\rho]$ is the ground-state electrostatic energy of $N$ point charges in an attractive background of density $\rho(\rv)$, including the background-background repulsion,
\begin{equation}\label{eq:EelDef}
	E_{el}[\rho]= \min_{\{\rv_1\dots\rv_N\}}\left\{\sum_{\substack{i,j=1 \\ j>i}}^{N}\frac{1}{|\rv_i-\rv_j|}-\sum_{i=1}^N v_h(\rv_i;[\rho])+U[\rho]\right\}.
\end{equation}
In other words, the $\lambda\to\infty$ limit of the MP adiabatic connection is a crystal bound by the ``positive'' charge density $\rho^{\rm HF}(\rv)$. 

The fact that $\alpha(\lambda)$ in Eq.~\eqref{eq:PsiT} must grow as $\lambda^{1/2}$ when $\lambda\to\infty$ has been found variationally in Ref.~\onlinecite{SeiGiaVucFabGor-JCP-2018}, by writing $\alpha=a\,\lambda^n$ and minimising the subleading term as $\lambda\to\infty$ with respect to $n$. In fact, by using the following notations
\begin{equation}\label{eq:Olambda}
\langle \hat{O} \rangle_\lambda=\left\langle\Psi_{\lambda}\left|\hat{O}\right|\Psi_{\lambda}\right\rangle     
\end{equation}
and
\begin{equation}\label{eq:Oexp}
\langle \hat{O}\rangle_{\lambda\rightarrow\infty}=O_{\infty}+\sum_{n=2}^\infty\lambda^{-\frac{n}{4}}\,O_{\frac{n}{4}}
\end{equation}
where $\hat{O}$ can be any operator independent of $\lambda$, we have that, with the trial wave function of Eq.~\eqref{eq:PsiT} and $\alpha(\lambda)=a\,\lambda^{1/2}$,
\begin{equation}
\langle \hat{K} \rangle_{\lambda\to\infty}= \lambda^{-1/2}K_{1/2}+O(\lambda^{-3/4}), 
\end{equation}
which shows that it is possible to make the expectation of $\hat{K}$ vanish at large $\lambda$ (although, of course $\Psi_\lambda^{T}$ will not provide in general the exact value of $K_{1/2}$). Since
\begin{equation}
	W_{c,\lambda}^{\rm HF}=\frac{d E_\lambda^{\rm HF}}{d\lambda}+U+E_x,
\end{equation}
with $E_\lambda^{\rm HF}$ the ground-state energy of Eq.~\eqref{eq:HlambdaHF}, 
$\langle \hat{K} \rangle_{\lambda}$ enters in the large-$\lambda$ expansion of $W_{c,\lambda}^{\rm HF}$ at the same order as the kinetic energy operator, whose expectation value diverges as $\lambda^{1/2}$ for large $\lambda$ (and thus, its derivative vanishes as $\lambda^{-1/2}$). Since the variational ansatz of Eq.~\eqref{eq:PsiT} provides the lowest possible expectation of $\hat{V}_{ee}-\hat{J}$, it also yields the exact $W^{\rm HF}_{c,\infty}$ in Eq.~\eqref{eq:HFSeidl}.\cite{SeiGiaVucFabGor-JCP-2018} 
The next leading order, however, is not exactly described by Eq.~\eqref{eq:PsiT} (even if we refine the ansatz with a normal modes analysis), and, as we will illustrate with the case of the H atom that is analytically soluble, $W^{\rm HF}_{\frac{1}{2}}$ has a different physics than its DFT counterpart,\cite{GorVigSei-JCTC-09} with the wave function of Eq.~\eqref{eq:PsiT} only providing a reasonable upper bound for it.

\section{The H atom: spin-polarized and unpolarized}\label{sec:H}

In this section we consider the hydrogen atom ($N=1$), both in the spin-polarized case (denoted here as H$[1,0]$), for which HF yields the exact ground-state energy and wave function (but not the exact spectrum), and in the spin-unpolarized case, with $1/2$ spin-up and $1/2$ spin-down electron (denoted as H$[\frac{1}{2},\frac{1}{2}]$), which appears locally in the stretched H$_2$ molecule treated in restricted HF, and it is often considered as a paradigmatic case for strong (static) correlation.\cite{CohMorYan-SCI-08,CohMorYan-JCP-08,CohMorYan-CR-12,Sav-CP-09,VucWagMirGor-JCTC-15} The two cases can be treated in a unified way by writing the hamiltonian of Eq.~\eqref{eq:HlambdaHF} as,
\begin{equation}\label{eq:HydrogenHF}
    \hat{H}_{\lambda}^{\rm HF}=\hat{T}+\hat{V}_{\rm ext}+(1-\lambda)\left(\hat{J}[\phi_s]-s\hat{K}[\phi_s]\right),
\end{equation}
where $s=1$ for  H$[1,0]$ and $s=1/2$ for H$[\frac{1}{2},\frac{1}{2}]$. The operators $\hat{J}$ and $\hat{K}$ are defined here in terms of the spatial HF orbital $\phi_s(\rv)$, with $\hat{J}$ being the Hartree local multiplicative operator,
\begin{equation}\label{eq:J}
    \hat{J}=\int d\textbf{r}'\frac{ \left|\phi_s(\textbf{r}')\right|^2}{\left|\textbf{r}-\textbf{r}'\right|}=v_{h}(\rv)
\end{equation}
and the action of $\hat{K}$ on a spatial wave function $\Psi(\rv)$ given by 
\begin{equation}\label{eq:K}
\bigl(\hat{K} \Psi\bigr)(\rv)=\phi_{s}(\textbf{r})\int d\textbf{r}'\frac{\phi_{s}^{*}(\textbf{r}')\Psi(\textbf{r}')}{\left|\textbf{r}-\textbf{r}'\right|}.
\end{equation}
These definitions in terms of spatial wave functions imply that for the H$[1,0]$ case we only search for minimising wave functions $\Psi_\lambda$ that have the same spin as the one at $\lambda=0$, i.e, that we forbid spin flip as $\lambda$ increases from 0 to $\infty$ (see Appendix~\ref{app:spinflipH}). For the  H$[\frac{1}{2},\frac{1}{2}]$ case the spin of the wave function $\Psi_\lambda$ does not matter.

The HF orbital $\phi_s(\rv)$ depends on $s$ and solves the non-linear problem at $\lambda=0$,
\begin{equation}\label{eq:phismin}
	\phi_s(\rv)=\text{arg}\min_{\phi}\langle\phi|\hat{T}+\hat{V}_{\rm ext}+\hat{J}[\phi]-s\hat{K}[\phi]|\phi\rangle.
\end{equation}
For $s=1$, the minimizer of Eq.~\eqref{eq:phismin}  will just be the hydrogen ground-state wave function, with radial part $\phi_{s=1}=2e^{-r}$, since the expectation of $\hat{J}[\phi]-\hat{K}[\phi]$ on $\phi$ is always zero, so that one ends up minimising $\hat{T}+\hat{V}_{\rm ext}$ alone. For $s=\frac{1}{2}$ the self-consistent HF solution (for which here we have used a basis of 10 STOs) gives a more diffuse orbital, since $\frac{1}{2}\hat{K}$ cannot fully remove the unphysical self-interaction of $\hat{J}$ anymore, which pushes the electron further from the nucleus. Notice that the SCF procedure is only needed to obtain the HF orbital ($\lambda=0$), while finding the wave function and the energy for all $\lambda>0$ is a simple linear eigenvalue problem, because $\hat{J}$ and $\hat{K}$ are fixed by the HF orbital $\phi_s$.

Since $\phi_s$ is spherically symmetric, performing the usual partial-wave expansion the hamiltonian \eqref{eq:HydrogenHF} becomes block-diagonal in each angular momentum channel $l$, each with energy
\begin{widetext}
\begin{align}\label{eq:HHFEnergy}
   & E_\lambda(l) = \min_u \biggl\{ \frac{1}{2}\int_{0}^{\infty}dr\, u'(r)^2  + \int_{0}^{\infty}dr\, u(r)^2 \left(\frac{l(l+1)}{2r^2}-\frac{Z}{r}+(1-\lambda)v_{h}(r)\right)
    \\ \nonumber & -\frac{2s}{2l+1}(1-\lambda)\int_{0}^{\infty}dr\, r^{-l} u(r)\phi_{s}(r)\int_{0}^{r}dr'\,r'^{l+1}\phi_{s}(r')u(r') \biggr\} , \qquad {\rm with}\; u(0)=0,\quad \int_0^\infty u(r)^2 dr=1
\end{align}
\end{widetext}
where $u(r)=r\psi_\lambda(r)$, and $\psi_\lambda(r)$ is the radial wave function. In Eq.~\eqref{eq:HHFEnergy} the radial HF orbital $\phi_s(r)$ is  normalised as $\int_0^\infty r^2\phi_s(r)^2 dr=1$, and we have used a general nuclear charge $Z$: while all the numerical computations are done at $Z=1$, the analytical derivation for the large-$\lambda$ asymptotics is carried out for a general $Z$.


\subsection{Computational Details}
We have computed $E_\lambda(l)$ and the minimizing $u(r)$ of Eq.~\eqref{eq:HHFEnergy}  for different channels $l$ in two different ways, because as $\lambda$ increases the energies for $l=0$ and $l=1$ can become very close, even crossing more than once in the $s=1$ case, and we wanted to be sure that our solver is accurate enough to capture this subtle feature.

The first method we have used is a simple variational minimisation using an STO basis set, 
\begin{equation}\label{eq:STO}
    \psi(r) = \sum_{n=1}^{10}c_{n}r^{n-1} e^{-a_{n}r}
\end{equation}
where we have also optimised the exponents $a_n$, although we have observed that setting all $a_n=1$ for $\lambda$ between 0 and 30 does not significantly change the energy. At $\lambda \gtrsim 30$, however, the wave function needs to contract to localize the electron in the minimum of $-v_h(r)$, and then we need to optimise the $a_n$ to obtain good energies.

The second method is the numerical solution on a grid of the Euler-Lagrange equation corresponding to the problem  \eqref{eq:HHFEnergy}, 
\begin{widetext}
\begin{align}\label{eq:SN2}
    \epsilon_{\lambda}u(r)&+\frac{1}{2}u''(r)-\frac{l(l+1)}{2r^2} u(r)=\left(\frac{-Z}{r}+(1-\lambda)v_{h}(r)\right)u(r)  \\\nonumber&-s\frac{(1-\lambda)}{2l+1}\phi_s(r)\left(r^{-l}\int_{0}^{r}dr'r'^{l+1}\phi_s(r')u(r')+r^{l+1}\int_{r}^{\infty}dr'r'^{-l}\phi_s(r')u(r')\right),
\end{align}
\end{widetext}
for which we have used the spectral renormalization (SR) method,\cite{AblMus-OL-05,AblMus-PRL-13,AblMus-NL-16} which was originally developed in the field of non-linear optics to find localized solitons. More recently, the SR method has been applied to converge the self-consistent Kohn-Sham equations with the functionals from the $\lambda\to\infty$ limit of the density-fixed DFT adiabatic connection.\cite{GroMusSeiGor-JPCM-20}
The SR variant we have used here starts from an initial $u^{(0)}$ to compute, via Eq.~\eqref{eq:HHFEnergy}, a first estimate of the eigenvalue $\epsilon_{\lambda}^{(0)}$. The next $u^{(1)}$ is computed from  \eqref{eq:SN2} by applying the inverse of the operator on the left-hand side to the right-hand side computed with $u^{(0)}$, and it is then normalised. The procedure is then repeated until convergence is reached. A main advantage of this method is that it does not depend on the initial guess and converges to a fixed point after only a few iterations. Although it has not been proven that the SR method always finds the global minimum (only a fixed point), experience shows that it actually always finds the ground state.\cite{AblMus-OL-05,AblMus-PRL-13,AblMus-NL-16} Nonetheless, here we compare the SR results to the variational basis-set expansion, and, indeed, we find that the SR method always converges to the lowest state, giving an energy slightly better (lower) than the STO one.

\subsection{Numerical results: $E_\lambda^{\rm HF}$ and $W_{c,\lambda}^{\rm HF}$ }
\subsubsection{The spin-polarized  case ($s=1$)}\label{sec:Hatomresults}
\begin{figure}
    \centering
    \includegraphics[width=0.4\textwidth]{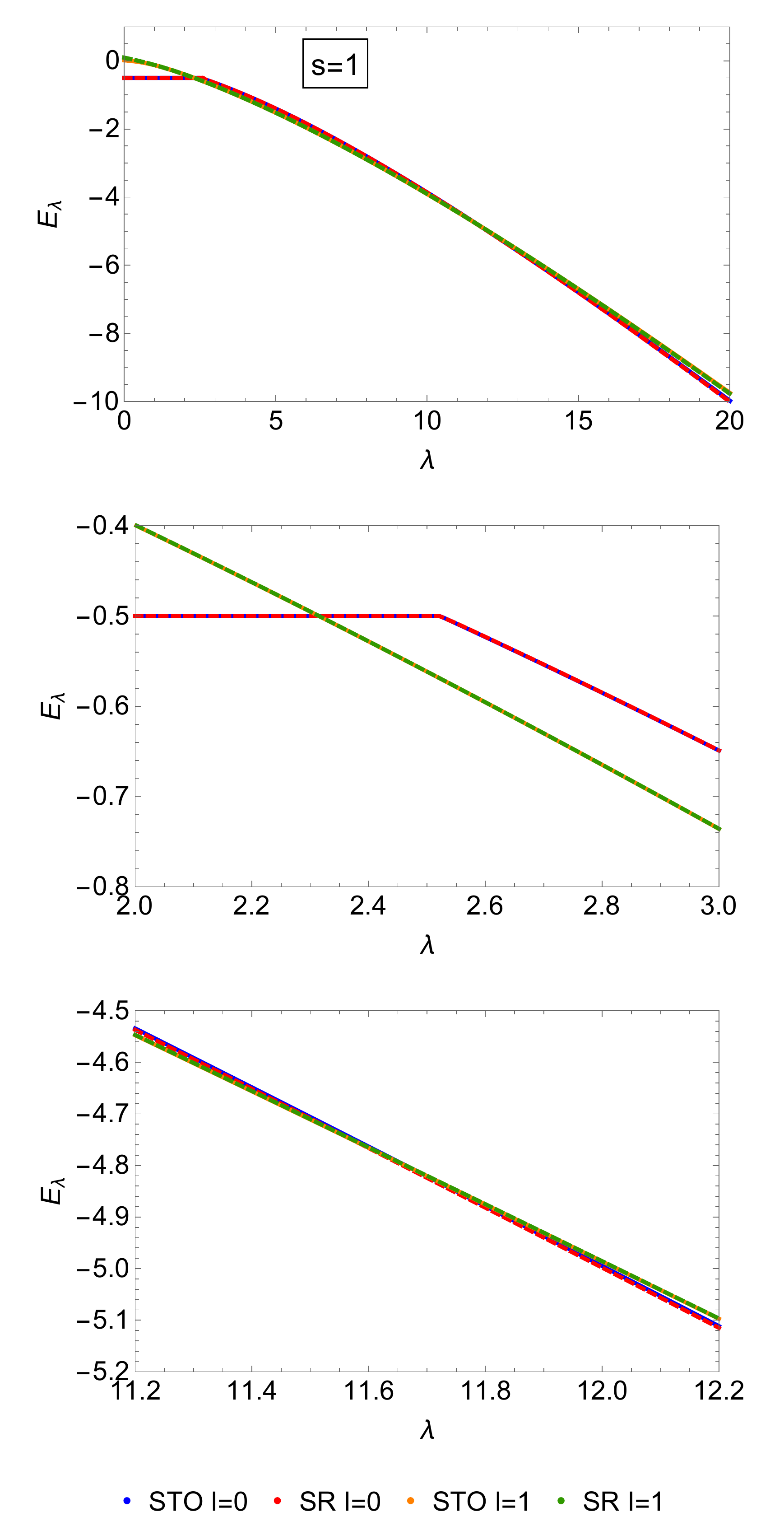}
    \caption{The lowest energy curves for H$[1,0]$ ($s=1$) computed with the STO expansion and the spectral renormalization (SR) method (upper panel). As $\lambda$ increases beyond 1, there are two crossings of states: first from the H atom ground state ($E_\lambda=-\frac{1}{2}$, $l=0$) to $l=1$ at $\lambda\approx 2.3$ (enlarged in the second panel), and then back to an $l=0$ channel at $\lambda\approx 11.6$ (enlarged in the third panel).} 
    \label{fig:ESTOSN1}
\end{figure}

The  H$[1,0]$ ($s=1$) system is trivial for $0\le \lambda \le 1$ (as in the DFT AC for all $\lambda\ge 0$): since $\hat{J}[\phi]-\hat{K}[\phi]$ is a positive definite operator, as long as $(1-\lambda)\ge 0$ in Eq.~\eqref{eq:HydrogenHF}, the best we can do is to make it vanish, which is achieved if $\Psi_\lambda=\phi_s$. Thus, as long as $\lambda \le 1$, the $\Psi_\lambda$ that minimises the hamiltonian of Eq.~\eqref{eq:HydrogenHF} will be the hydrogenic ground state. As soon as $\lambda>1$, however, the situation changes, since it starts to be variationally convenient to make the expectation of $\hat{J}[\phi]-\hat{K}[\phi]$ different from zero. Interestingly, this happens at a $\lambda$ quite larger than 1, $\lambda\approx 2.3$ ($\lambda=2.3144$ with the STO expansion and $\lambda=2.3142$ with the SR method), with the ground state switching to the $l=1$ channel, as shown in Fig.~\ref{fig:ESTOSN1}.  Around $\lambda\approx 11.6$ ($\lambda=11.68$ with the STO expansion and $\lambda=11.55$ with the SR method), there is a second crossing of states, in which the $l=0$ channel becomes again the lowest. The other $l$ channels give energies much higher at all $\lambda\ge 0$. 

The $l=0$ channel remains the lowest for all $\lambda\gtrsim 11.6$: in Sec.~\ref{sec:order1/2} we will also provide the analytic solution for $\Psi_{\lambda\to\infty}$, which provides the exact $W_{c,\lambda}^{\rm HF}$ up (and including) orders $\lambda^{-3/4}$.

\begin{figure}
    \centering
    \includegraphics[width=0.4\textwidth]{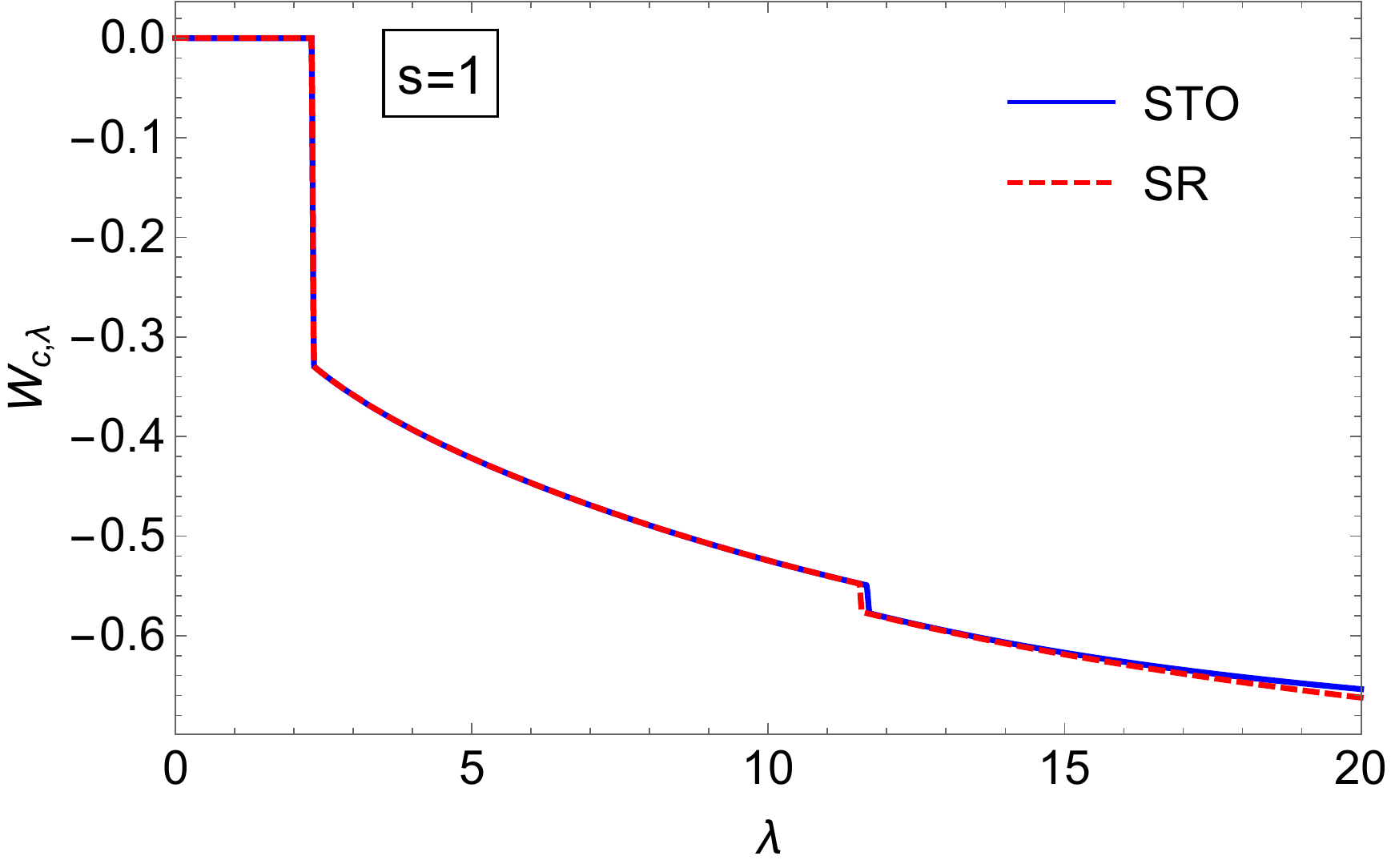}
    \caption{$W_{c,\lambda}^{\rm HF}$ for   H$[1,0]$ ($s=1$) computed with the STO expansion and the spectral renormalization method (SR) between $\lambda=0$ and $\lambda=20$. The two jumps appear at the two crossings of states of Fig.~\ref{fig:ESTOSN1}. } 
    \label{fig:WSTOSN1}
\end{figure}
In Fig.~\ref{fig:WSTOSN1} we also report the corresponding $W_{c,\lambda}^{\rm HF}$, which obviously has jumps at the $\lambda$ values when we have a crossing of states. These crossings of states are expected to occur more often in the MP AC than in the density-fixed DFT AC, as in the latter the density constraint enforces many symmetries. This clearly makes it more difficult to build interpolations, which would somehow be an average over the discontinuites. We should also remark that the H$[1,0]$ case is particularly pathological for the MP AC (see also the discussion in Sec.~\ref{sec:SIornotSI}) and should not be considered as very representative for the general case.

\subsubsection{The spin-unpolarized case ($s=1/2$)}
\begin{figure}[htp]
    \centering
    \includegraphics[width=0.4\textwidth]{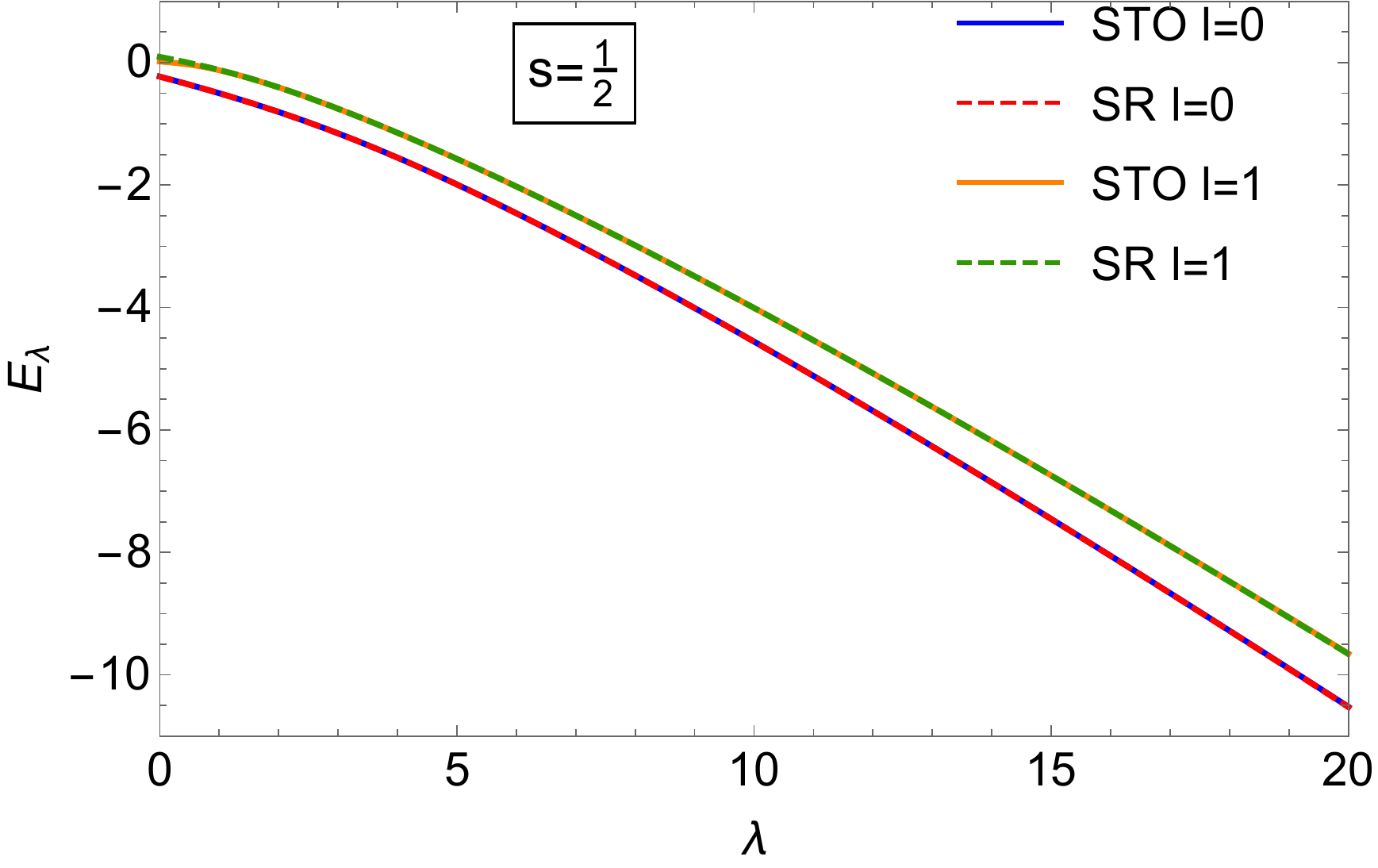}
    \caption{The lowest energy curves for H$[\frac{1}{2},\frac{1}{2}]$ ($s=\frac{1}{2}$) computed with the STO expansion and the spectral renormalization (SR) method for both the $l=0$ and $l=1$ channel.} 
    \label{fig:ESTOSN12}
\end{figure}

\begin{figure}[htp]
    \centering
    \includegraphics[width=0.4\textwidth]{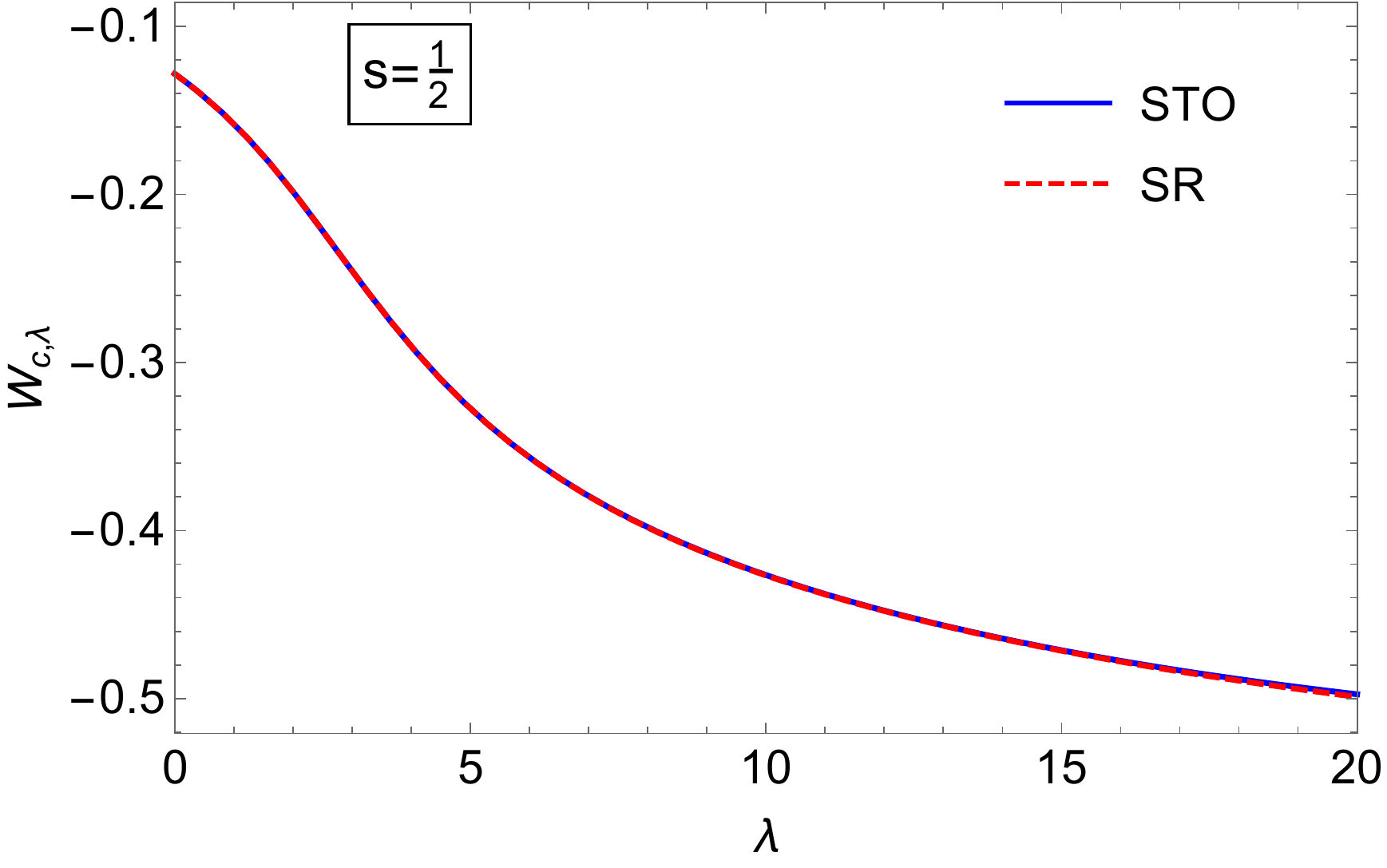}
    \caption{$W_{c,\lambda}^{\rm HF}$ for H$[\frac{1}{2},\frac{1}{2}]$ ($s=\frac{1}{2}$) computed with the STO expansion and the spectral renormalization (SR) method  between $\lambda=0$ and $\lambda=20$.} 
    \label{fig:WSTOSN12}
\end{figure}

In the H$[\frac{1}{2},\frac{1}{2}]$ case, the $l=0$ channel turns out to be always the lowest in energy, as shown in Fig.~\ref{fig:ESTOSN12}. The absence of crossing of states means that $W_{c,\lambda}^{\rm HF}$, reported in Fig.~\ref{fig:WSTOSN12}, is now a smooth function of $\lambda$. We can also see that $W_{c,\lambda}^{\rm HF}$ has a peculiar shape, changing from a concave to a convex curve, which is expected to be a general feature of the MP adiabatic connection integrand. In fact,
 MP2 typically yields correlation energies that are too small in absolute value (i.e., too high), implying that at $\lambda=0$ the tangent to $W_{c,\lambda}^{\rm HF}$ lies above the curve. For example, in Refs.~\onlinecite{Per-JCP-18,VucFabGorBur-JCTC-20},
 $W_{c,\lambda}^{\rm HF}$ for the He isoelectronic series and for the H$_2$ molecule has been computed  for $\lambda$ between 0 and 1, where it has been found to be concave in this $\lambda$ range. However, since it has been proven\cite{SeiGiaVucFabGor-JCP-2018} that when $\lambda\to\infty$ $W_{c,\lambda}^{\rm HF}$ tends to a finite value, at some finite $\lambda$ the curve needs to become convex.
 
Notice that $W_{c,\lambda=0}^{\rm HF}$ is not zero, but equal to $-\frac{1}{2}U$. This is because the H$[\frac{1}{2},\frac{1}{2}]$ system we are considering is a subsystem, namely a H atom inside an infinitely stretched H$_2$ molecule treated in restricted HF. If we consider the whole system (the molecule), as we will do in Sec.~\ref{sec:H2}, then $W_{c,\lambda=0}^{\rm HF}=0$, and as the internuclear distance $R$ becomes very large, the slope of $W_{c,\lambda}^{\rm HF}$  at $\lambda=0$ (MP2) tends to $-\infty$. In the $R\to\infty$ limit, the adiabatic connection curve for the H$_2$ molecule ``jumps'' to twice the curve of our Fig.~\ref{fig:WSTOSN12}, as will be shown in Fig.~\ref{fig:H2}.

\subsection{Numerical Results: The Minimizing Wave Functions}
\subsubsection{The spin-polarized case ($s=1$)}
\begin{figure}[htp]
    \centering
    \includegraphics[width=0.4\textwidth]{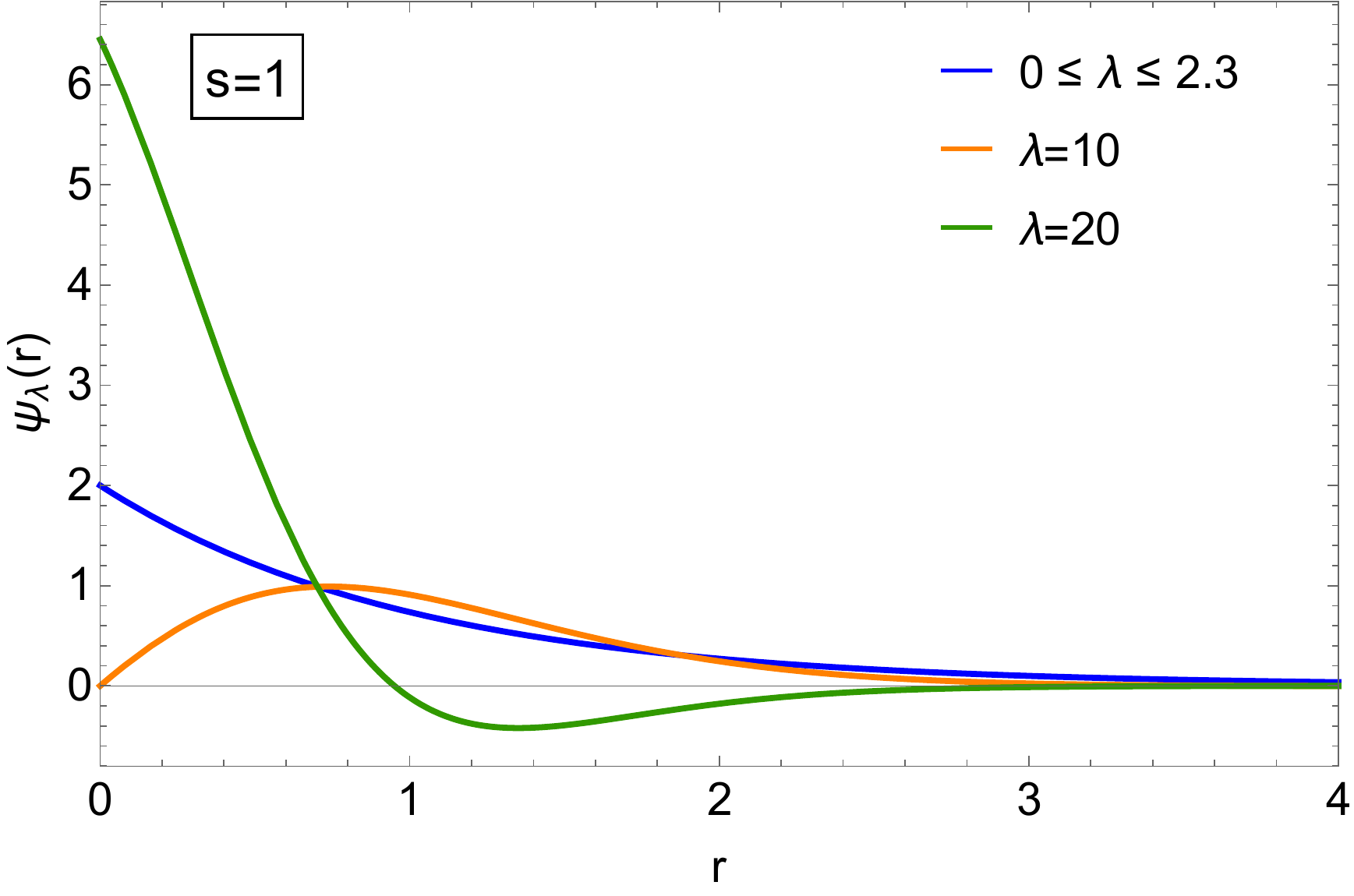}
    \caption{The radial wave functions $\psi_\lambda(r)=u(r)/r$, where $u(r)$ minimizes Eq.~\eqref{eq:HHFEnergy} for H$[1,0]$ ($s=1$) in the $l$-channel with the lowest energy, which is $l=0$ for $0\le\lambda\lesssim 2.3$, then $l=1$ for $2.3\lesssim \lambda \lesssim 11.6$, and again $l=0$ for larger $\lambda$, computed with the spectral renormalization (SR) method.} 
    \label{fig:WFSTOSN1}
\end{figure}

As said, for the $s=1$ system the minimizing wave function is just the hydrogenic 1$s$ orbital for $0\le\lambda\lesssim 2.3$, switching to a radial $l=1$ wave function at $\lambda\approx 2.3$, which develops a radial node associated to a single oscillation with a small amplitude. This oscillation might be present already at the crossing of states but for very large $r$ and with a very small amplitude and becomes more evident as $\lambda$ increases. At the second crossing ($\lambda\approx 11.6$), when the lowest energy state becomes again $l=0$, we have a function which also has a radial node, as shown in Fig.~\ref{fig:WFSTOSN1}. This single radial node remains present as $\lambda$ increases, also in the limit $\lambda\to\infty$, which can be computed analytically and will be presented and discussed in Sec.~\ref{sec:largelambdaH}.

\subsubsection{The spin-unpolarized case ($s=1/2$)}
\begin{figure}[htp]
    \centering
    \includegraphics[width=0.4\textwidth]{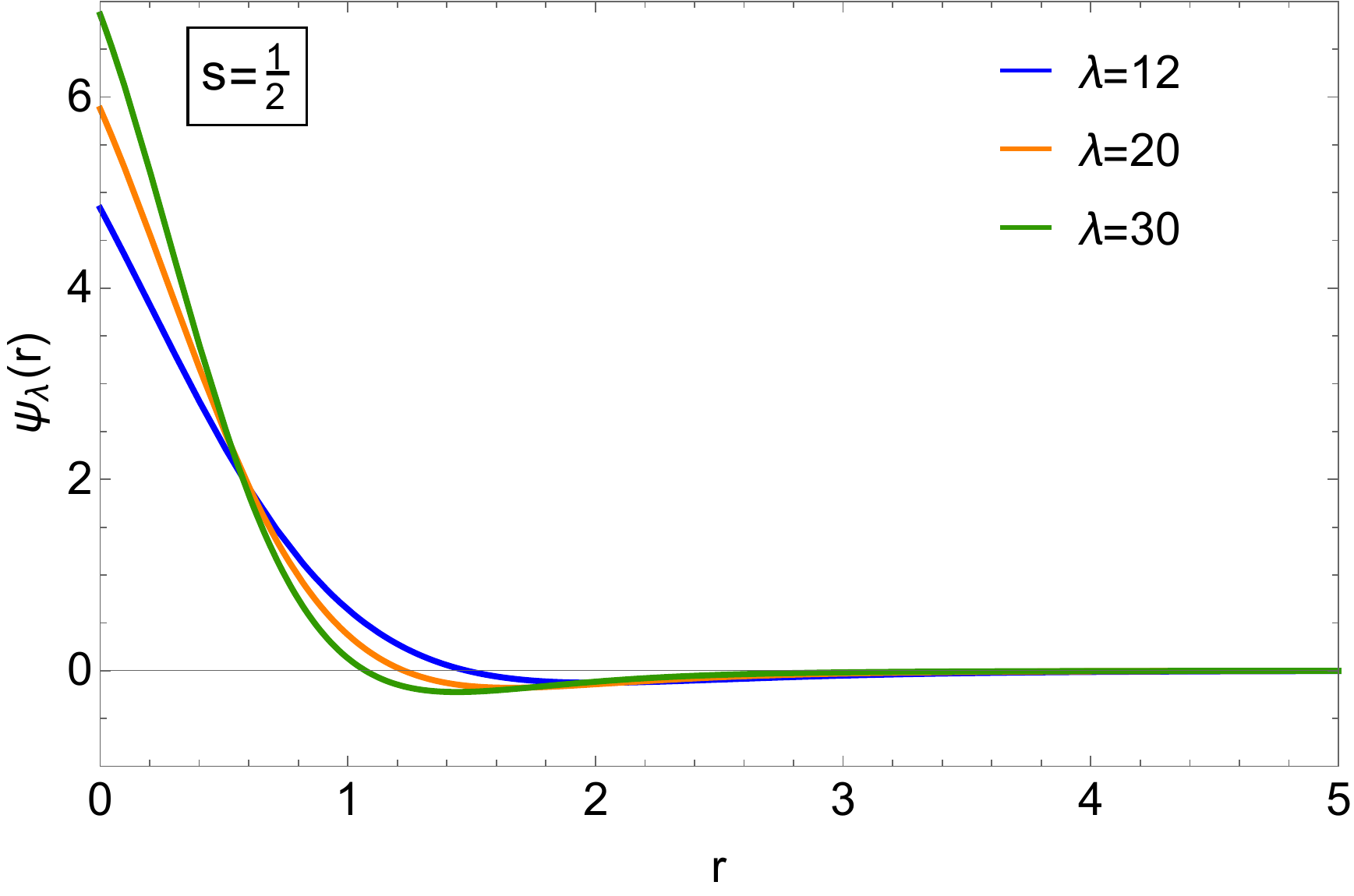}
	\includegraphics[width=0.4\textwidth]{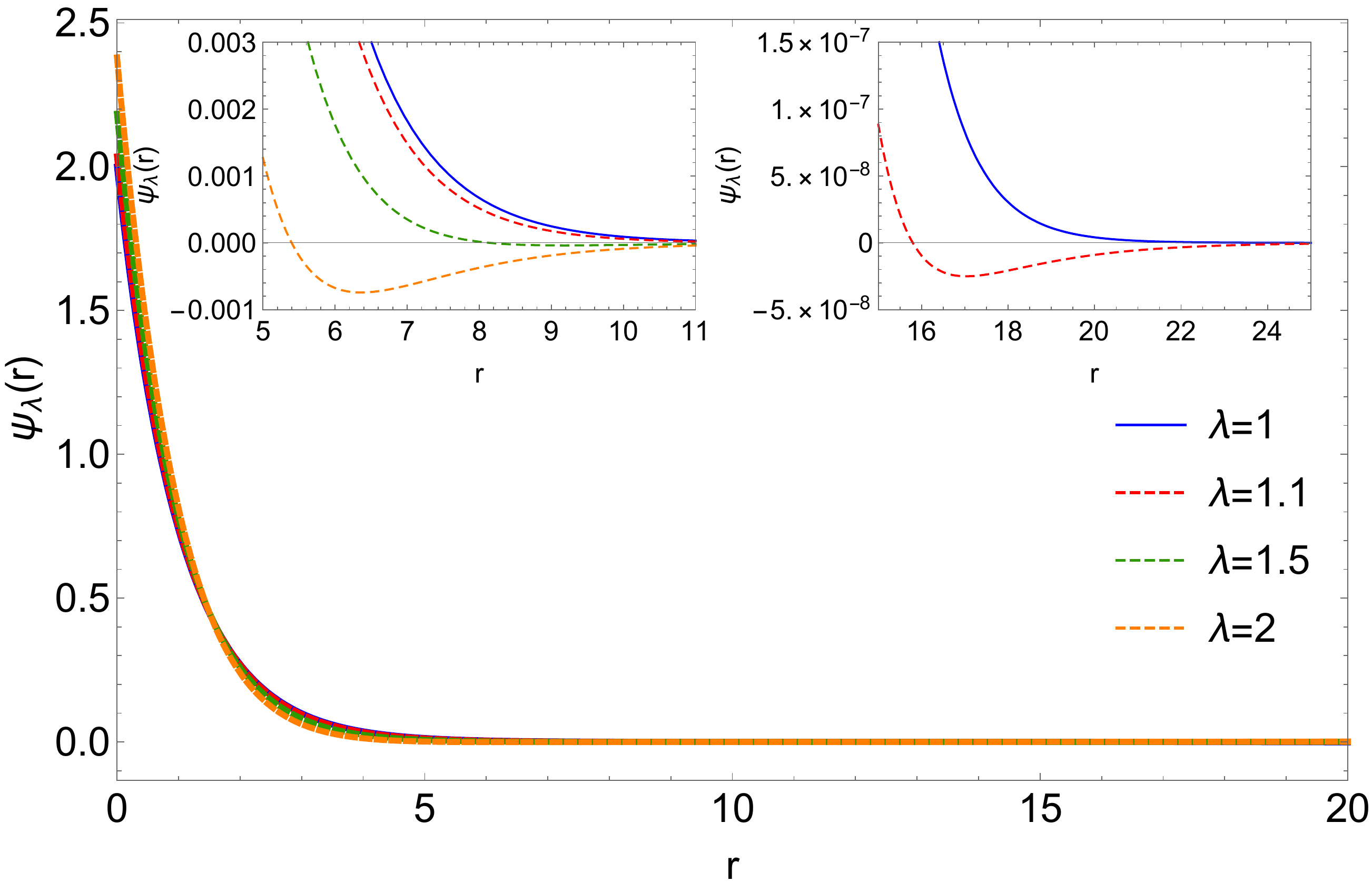}
    \caption{The radial wave functions $\psi_\lambda(r)=u(r)/r$, where $u(r)$ minimizes Eq.~\eqref{eq:HHFEnergy} with $l=0$,   for H$[\frac{1}{2},\frac{1}{2}]$ ($s=\frac{1}{2}$), computed with the spectral renormalization (SR) method.
	The appearance of the radial node as soon as $\lambda>1$ is illustrated in the second panel.} 
    \label{fig:WFSTOSN12}
\end{figure}

In the $s=\frac{1}{2}$ system the ground-state wave function switches from a nodeless  function for  $0\le\lambda\le 1$ to one with a radial node, apparently as soon as $\lambda>1$. In fact,  we observe a node already at $\lambda=1.1$, as shown in Fig.~\ref{fig:WFSTOSN12}, where we see that the node appears first at large $r$, and then moves inwards (towards smaller $r$) as $\lambda$ increases. Also, the single oscillation associated with the node starts with a very tiny amplitude as soon as $\lambda>1$ and increases in amplitude as $\lambda$ grows. When $\lambda\to\infty$, we will see that the wave function contracts,  keeping one node, although for the  $s=\frac{1}{2}$ case we could not find an analytic solution for large $\lambda$. The presence of the node at large $\lambda$ for both $s=1$ and $s=\frac{1}{2}$ will be explained in Sec.~\ref{sec:largelambdaH}.

\subsection{The large-$\lambda$ limit}\label{sec:largelambdaH}

\subsubsection{Scaling and large-$\lambda$ expansion}
In this derivation we try to be as general as possible and keep track of $l$, $s$ and the nuclear charge $Z$, starting from the Euler-Lagrange equation \eqref{eq:SN2} for $u(r)=r\psi_\lambda(r)$, where $\psi_\lambda(r)$ is the radial part of $\Psi_\lambda(\rv)$.
As shown in Ref.~\onlinecite{SeiGiaVucFabGor-JCP-2018}, when $\lambda\to\infty$ we end up minimising the expectation of a classical potential energy given by $\hat{V}_{ee}-\hat{J}$. In this case, with $N=1$ electrons, we need to minimize the expectation of $-v_h(r)$ alone. The square of the wave function will then tend asymptotically to a Dirac delta function centered in the minimum of $-v_h(r)$, which is at the nucleus. At the next leading order, we might expect, as in DFT,\cite{GorVigSei-JCTC-09} zero-point oscillations around this minimum, although, as we will see, the presence of the operator $\hat{K}$ will alter the wave function at orders $\lambda^{1/2}$, introducing a node. Nonetheless, the scaling at large $\lambda$ remains the same\cite{SeiGiaVucFabGor-JCP-2018} as in the DFT case (at least in 3D), with the relevant scaled coordinate being \cite{GorVigSei-JCTC-09}
\begin{equation}\label{eq:t}
 t=\lambda^{\frac{1}{4}}r.
\end{equation}
When we rewrite Eq.~\eqref{eq:SN2} in terms of $t$, we will have that, as $\lambda\to\infty$, both $\phi_s(\lambda^{-\frac{1}{4}}t)$ and $v_h(\lambda^{-\frac{1}{4}}t)$ can be expanded around the origin, 
\begin{align} \phi_s(\lambda^{-\frac{1}{4}}t)& =\phi_s(0)+\phi_s'(0)\lambda^{-\frac{1}{4}}t+\ldots \nonumber \\&=\phi_s(0)(1-Z\lambda^{-\frac{1}{4}}t)+\ldots \label{eq:ScaledHFO}
\end{align}
where we have used the cusp condition, $\phi_s'(0)=-Z\phi_s(0)$, in the last equation. Similarly, for $v_h$ we have 
\begin{equation}\label{HFscaled}
    v_{h}(\lambda^{-\frac{1}{4}}t)=v_{h}(0)-\frac{1}{6}\phi_s(0)^2 \lambda^{-\frac{1}{2}}t^2 + \frac{Z}{6}\phi_s(0)^2 \lambda^{-\frac{3}{4}}t^3+\ldots.
\end{equation}
Inserting these expansions in Eq.~\eqref{eq:SN2} we can collect the different orders for large $\lambda$  
\begin{widetext}
\begin{align}\label{eq:HFExpansion}
    &\nonumber-\lambda\, v_{h}(0)\,u(t)+\lambda^{\frac{1}{2}}\left[-\frac{1}{2}u''(t)+\frac{l(l+1)}{2t^2}u(t)+\frac{1}{6}\phi_s(0)^2 t^2u(t)+\frac{s\,\phi_s(0)^2}{2l+1}\left(t^{-l}\int_{0}^{t}dt't'^{l+1}u(t')+t^{l+1}\int_{t}^{\infty}dt't'^{-l}u(t')\right)\right]\\\nonumber&-Z\lambda^{\frac{1}{4}}\biggl[\frac{u(t)}{t}+ \frac{1}{6}\phi_s(0)^2 t^3 u(t)+\frac{s\,\phi_s(0)^2}{2l+1}\biggl(t^{1-l}\int_{0}^{t}dt't'^{l+1}u(t')+t^{l+2}\int_{t}^{\infty}dt't'^{-l}u(t')+t^{-l}\int_{0}^{t}dt't'^{l+2}u(t')\\&+t^{l+1}\int_{t}^{\infty}dt't'^{1-l}u(t')\biggr)\biggr]+O(\lambda^0)=\left(\lambda\, \epsilon_\infty+\lambda^{\frac{1}{2}}\,\epsilon_{\frac{1}{2}}+\lambda^{\frac{1}{4}}\,\epsilon_{\frac{1}{4}}+O(\lambda^0)\right)u(t),
\end{align}
\end{widetext}
where we have also carried out the same large-$\lambda$ expansion for the eigenvalue $\epsilon_\lambda$, and we have improperly used the same symbol $u(t)$ for the function $u(\lambda^{-1/4}t)$. We then immediately see that, as predicted, the leading term is not affected by $\hat{K}$ and it is given by the minimum of $-v_h(r)$, which is at the nucleus,
\begin{equation}
	\epsilon_\infty=-v_{h}(0).
\end{equation}
Since this leading term is independent of $l$, it is the order $\lambda^{1/2}$ that  determines which $l$ channel will be the lowest in the large $\lambda$ limit. 


\subsubsection{The order $\lambda^{1/2}$}\label{sec:order1/2}
From Eq.~\eqref{eq:HFExpansion} we can directly read the pseudo eigenvalue equation for the order $\lambda^{1/2}$, which, by defining
\begin{equation}
	p=\sqrt{\phi_s(0)}\,t=\lambda^{\frac{1}{4}}\sqrt{\phi_s(0)}\,r, \qquad \Tilde{\epsilon}_{\frac{1}{2}}=\frac{\epsilon_{\frac{1}{2}}}{\phi_s(0)},
\end{equation}
can be further simplified into
\begin{widetext}
\begin{equation}\label{eq:HFlb12}
    -\frac{1}{2}u_{\frac{1}{2}}''(p)+\frac{l(l+1)}{2p^2}u_{\frac{1}{2}}(p)+\frac{1}{6}p^2 u_{\frac{1}{2}}(p) + \frac{s}{2l+1}\left(p^{-l}\int_{0}^{p}dq~q^{l+1}u_{\frac{1}{2}}(q)+p^{l+1}\int_{p}^{\infty}dq~ q^{-l}u_{\frac{1}{2}}(q)\right)=\Tilde{\epsilon}_{\frac{1}{2}}u_{\frac{1}{2}}(p),
\end{equation}
\end{widetext}
which depends only on $s$ and $l$. This equation turns out to have simple analytical solutions for certain pairs of $s$ and $l$, one of them being $s=1,l=0$. The other analytical solutions seem to be all for $s>1$ (e.g., $s=5/3,l=1$), which are not relevant for our problem. The simple analytical solutions are finite linear combinations of the 3D isotropic harmonic oscillator (IHO) eigenfunctions for the problem with $s=0$ in Eq.~\eqref{eq:HFlb12},
\begin{align}\label{eq:HFLargeLB}
 u_{\frac{1}{2}}(p) & = p\,\sum_n c_n \xi_{n}(p) \\
 \xi_{n}(p) & = \mathcal{N}\,\exp \left(-\frac{p^2}{2
   \sqrt{3}}\right) p^l L_n^{l+\frac{1}{2}}\left(\frac{p^2}{\sqrt{3}}\right) \nonumber \\
   \mathcal{N} & = \sqrt{\frac{\sqrt{\frac{1}{4 \pi  \sqrt{3}^3}} \left(2^{n+2 l+3} n!\right) \left(\frac{1}{2 \sqrt{3}}\right)^l}{(2 n+2 l+1)\text{!!}}} \nonumber 
\end{align}
where $L_{n}^{l+\frac{1}{2}}$ are the generalised Laguerre polynomials and the eigenvalues for the IHO are $(2n+l+\frac{3}{2})\,\omega$, with $\omega=1/\sqrt{3}$ in our case, and $n=0,1,2,\dots$.

\begin{figure}
    \centering
    \includegraphics[width=0.4\textwidth]{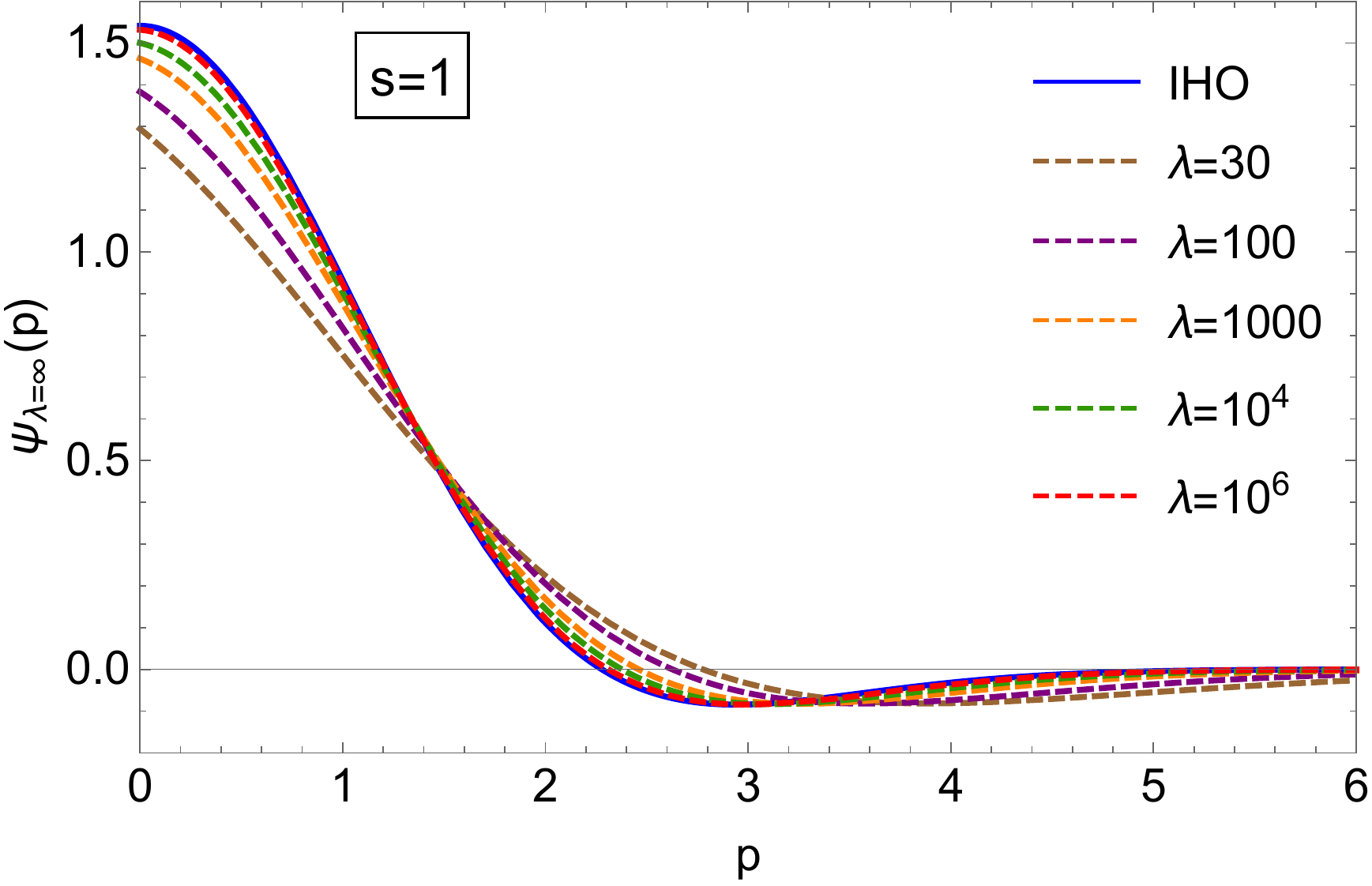}
    \caption{The analytical solution for $s=1,l=0$ for the order $\lambda^{1/2}$, given exactly by two IHO states, Eq.~\eqref{eq:HFs1sol}, compared to the spectral renormalization (SR) wave function for the full $\lambda$-dependent equation~\eqref{eq:SN2} at various $\lambda$, in the scaled coordinate $p=\lambda^{\frac{1}{4}}\sqrt{\phi_s(0)}\,r$.} 
    \label{fig:uopts1}
\end{figure}

For $s=1,l=0$ the analytic solution for $\psi_{\lambda\rightarrow\infty}^{\text{HF}}(p)=u_{\frac{1}{2}}(p)/p$ is
\begin{align}\label{eq:HFs1sol}
   \psi_{\lambda\rightarrow\infty}^{\text{HF}}(p) & =\frac{4 e^{-\frac{p^2}{2 \sqrt{3}}} \left(9-\sqrt{3} p^2\right)}{3^{15/8} \sqrt{5} \sqrt[4]{\pi }}   \\
   \Tilde{\epsilon}_{\frac{1}{2}} & =\frac{7}{2\sqrt{3}}\approx 2.02073, \quad (s=1,l=0), \nonumber
\end{align}
which is a linear combination of the ground state and the first excited state of the 3D IHO. In Fig.~\ref{fig:uopts1} we compare Eq.~\eqref{eq:HFs1sol} to the scaled wave function at $\lambda$ larger and larger obtained from the SR solution of the full $\lambda$-dependent equation~\eqref{eq:SN2}, finding perfect agreement for $\lambda=10^6$. 

\begin{table}
\caption{The value of $\tilde{\epsilon}_{\frac{1}{2}}$ for the $s=1$ and $s=\frac{1}{2}$ systems, for both the $l=0$ and $l=1$ channels using the 3D isotropic harmonic oscillator (IHO) basis set and the spectral renormalization (SR) method to solve Eq.~\eqref{eq:HFlb12}. For $s=1,l=0$, two IHO states solve Eq.~\eqref{eq:HFlb12} exactly, see Eq.~\eqref{eq:HFs1sol}. For the other cases we have used 11 IHO states, except for $s=\frac{1}{2},l=0$ for which we have used 21 IHO states. }\label{tb:Channels}
\begin{tabular}{cl|ll}
\multicolumn{2}{l|}{}                                   & $l=0$    & $l=1$  \\ \hline
\multirow{2}{*}{$s=1$}                              & IHO  & 2.0207 &  2.2357 \\
                                                  & SR  & 2.0210 & 2.2362   \\
												  \hline 
\multirow{2}{*}{$s=\frac{1}{2}$}                  & IHO  & 1.6185  & 1.9005  \\
                                                  & SR  & 1.6192  & 1.9007  
\end{tabular}
\end{table}

We thus see that the radial node observed at large but finite $\lambda$ persists in the $\lambda\to\infty$ limit, and it is due to the operator $\hat{K}$, which makes the asymptotic wave function different than the simple IHO ground state of the variational ansatz of Eq.~\eqref{eq:PsiT}. In the $s=1$ case, $\hat{K}$ simply mixes in the first IHO excited state. 
The reason why $\hat{K}$ must introduce a radial node in the $l=0$ channel can be understood by looking at Eq.~\eqref{eq:HFlb12}, in which the term due to $\hat{K}$ reads
\begin{equation}\label{eq:HFKnode}
   s\left(\int_{0}^{p}dq\,q\, u_{\frac{1}{2}}(q)+p\int_{p}^{\infty}dq\,u_{\frac{1}{2}}(q)\right).
\end{equation}
This term must vanish when $p\to\infty$ for a localised solution. This happens automatically for the second term of Eq.~\eqref{eq:HFKnode} above, but not for the first one. Thus, any localised solution of Eq.~\eqref{eq:HFlb12} must satisfy the additional constraint
\begin{equation}\label{eq:constr}
   \int_{0}^{\infty}dq\,q\, u_{\frac{1}{2}}(q)=0,
\end{equation}
which requires at least one radial node. The radial node also appears in the other cases ($l > 0$ and the $s = \frac{1}{2}$ system) for exactly the same reason.

\begin{figure}[htp]
    \centering
    \includegraphics[width=0.4\textwidth]{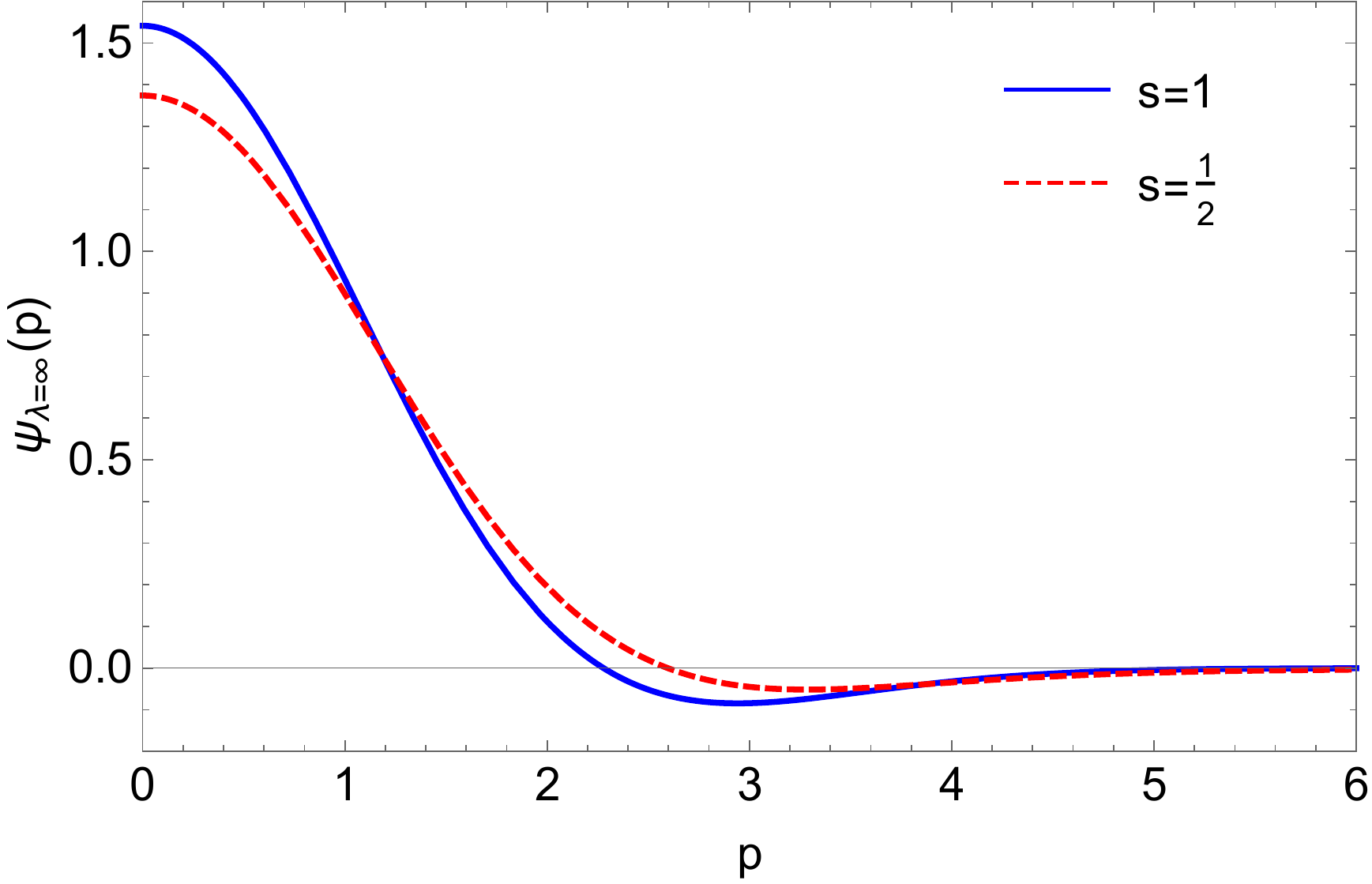}
    \caption{The solution $u_\frac{1}{2}(p)/p$ of Eq.~\eqref{eq:HFlb12} for $s=\frac{1}{2}$ obtained with 21 IHO states, compared to the solution for the $s=1$ case.  } 
    \label{fig:asyWFs}
\end{figure}

For $s=1/2$ or for $s=1$ and $l>0$, we can still use the IHO wave functions a finite basis approximation for $u_{\frac{1}{2}}$, observing a  reasonably fast convergence for the energy. In table~\ref{tb:Channels} we show the results for $\tilde{\epsilon}_{\frac{1}{2}}$ for both $s=1$ and $s=\frac{1}{2}$ in the $l=0$ and $l=1$ channels. We see that the $l=0$ channel remains the lowest as $\lambda\to\infty$ in both cases. The asymptotic wave function for $s=\frac{1}{2}$ has a shape similar to the one for $s=1$, as shown in Fig.~\ref{fig:asyWFs}, where it has been computed with 21 IHO basis functions.

We can compare this exact (or very accurate) asymptotic solutions for $s=1$ and $s=\frac{1}{2}$ with the variational ansatz of Eq.~\eqref{eq:PsiT}: we see that the exact (or accurate) wave functions have the same functional form of Eq.~\eqref{eq:PsiT}, with a localised function of the scaled variable $t=\lambda^{1/4}\,r$ centered at the minimum of $-v_h(r)$, whose square tends to a Dirac $\delta$ function  when $\lambda\to\infty$. The leading term of order $\lambda$ is then the same in both cases, as predicted. (This value does not depend on the particular representation we choose for the $\delta$ function, which could be, for example, any finite linear combination of IHO wave functions). The next leading term of order $\lambda^{1/2}$, however, selects the precise representation of the $\delta$ function, which is different than the simple IHO ground-state used in Eq.~\eqref{eq:PsiT}, mixing in the excited states.
The ground-state IHO of Eq.~\eqref{eq:PsiT}, with the $\omega$ optimized variationally, gives an upper bound for $\tilde{\epsilon}_{\frac{1}{2}}$, equal to $\frac{1}{2}\sqrt{3\,(1+8s)}$, corresponding to $2.5981$ for $s=1$ (compared to the exact $2.0207$) and $1.9365$ for $s=\frac{1}{2}$ (compared to the accurate value $1.6185$).

\subsubsection{The order $\lambda^\frac{1}{4}$}
If we subtract from both sides of Eq.~\eqref{eq:HFExpansion} the constant leading term of order $\lambda$ and divide everything by $\lambda^{1/2}$, we obtain for the operators on the left-hand side a perturbation expansion of the kind $\hat{H}_{\frac{1}{2}}+\lambda^{-1/4}\,\hat{H}_{\frac{1}{4}}+O(\lambda^{-1/2})$,
which implies that $\epsilon_{\frac{1}{4}}$ is exactly given by the first-order perturbation term $\langle u_{\frac{1}{2}}|\hat{H}_{\frac{1}{4}}|u_{\frac{1}{2}}\rangle$, i.e., $\epsilon_{\frac{1}{4}}=Z\,\sqrt{\phi_s(0)}\, \tilde{\epsilon}_{\frac{1}{4}}$, with
\begin{widetext}
\begin{align}
	\tilde{\epsilon}_{\frac{1}{4}}=-\left\{\int_0^\infty dp\, \left(\frac{1}{p}+\frac{p^3}{6}\right) u_{\frac{1}{2}}(p)^2+2s\,\left[\int_0^\infty dp\, p\, u_{\frac{1}{2}}(p)\int_0^p dq\, q\, u_{\frac{1}{2}}(q)+\int_0^\infty dp\, u_{\frac{1}{2}}(p)\int_0^p dq\, q^2\, u_{\frac{1}{2}}(q)\right]
	\right\},
\end{align}
\end{widetext}
where we have directly considered $l=0$ only, since we are interested in the ground-state energy. We then find
\begin{align}
	\tilde{\epsilon}_{\frac{1}{4}} & = -\frac{112}{15\times 3^{1/4}\sqrt{\pi}}\approx -3.2009\qquad & (s=1) \\
	\tilde{\epsilon}_{\frac{1}{4}} & = -2.70306  & (s=\tfrac{1}{2}), \label{eq:eps14s12}
\end{align}
where the $s=\frac{1}{2}$ value has been obtained with 21 IHO basis functions (with the SR method we get $-2.69993$).

\subsubsection{The large-$\lambda$ expansion of $W_{c,\lambda}^{\rm HF}$ }
Putting everything together, with the radial HF orbital $\phi_s(r)=\sqrt{4\pi\rho^{\rm HF}(r)}$, we find that $W_{c,\lambda}^{\rm HF}$ for the H atom at large $\lambda$ has the expansion
\begin{align}\label{eq:HFnonSeidl}
    W^{\rm HF}_{c,\lambda\rightarrow\infty} & = W^{\rm HF}_{c,\infty} + \frac{W^{\rm HF}_{\frac{1}{2}}}{\sqrt{\lambda}}+\frac{W^{\rm HF}_{\frac{3}{4}}}{\lambda^{\frac{3}{4}}}+\dots \\
	W^{\rm HF}_{c,\infty} & = -v_h(0)+(1-s)\,U \\
	W^{\rm HF}_{\frac{1}{2}} & = \tilde{\epsilon}_{\frac{1}{2}}\frac{\sqrt{4\pi}}{2}\sqrt{\rho^{\rm HF}(0)} \\
	W^{\rm HF}_{\frac{3}{4}} & = Z\,\tilde{\epsilon}_{\frac{1}{4}}\frac{\sqrt[4]{4\pi}}{4}\sqrt[4]{\rho^{\rm HF}(0)}\label{eq:lambda1/4}
\end{align}
The presence of the order $\lambda^{-3/4}$ is interesting, because this term is zero in the large $\lambda$-expansion of the DFT adiabatic connection.\cite{GorVigSei-JCTC-09} We see that here this order is non-zero because the position  $\rv^{\rm min}$ is at the nucleus, which makes (i) the external potential expectation value diverge as $\lambda^{1/4}$ and (ii) the third-order expansion of $-v_h(r)$ and the first-order expansion of the HF orbital around $\rv^{\rm min}$, which would normally have zero expectation on a spherically symmetric function around $\rv^{\rm min}$, be non-zero because of the cusp. In Sec.~\ref{sec:HFdensfunc} we generalize Eqs.~\eqref{eq:HFnonSeidl}-\eqref{eq:lambda1/4} to the closed-shell many-electron case, for which we can still expect that, in most cases, for each atom one of the $\rv^{\rm min}_i$ is at the nucleus (with exceptions, of course). In the uniform electron gas case, analyzed in Sec.~\ref{sec:UEG}, this term is zero, as the nuclear charge is ``smeared'' into a continuum background.

\begin{figure}
    \centering
    \includegraphics[width=0.4\textwidth]{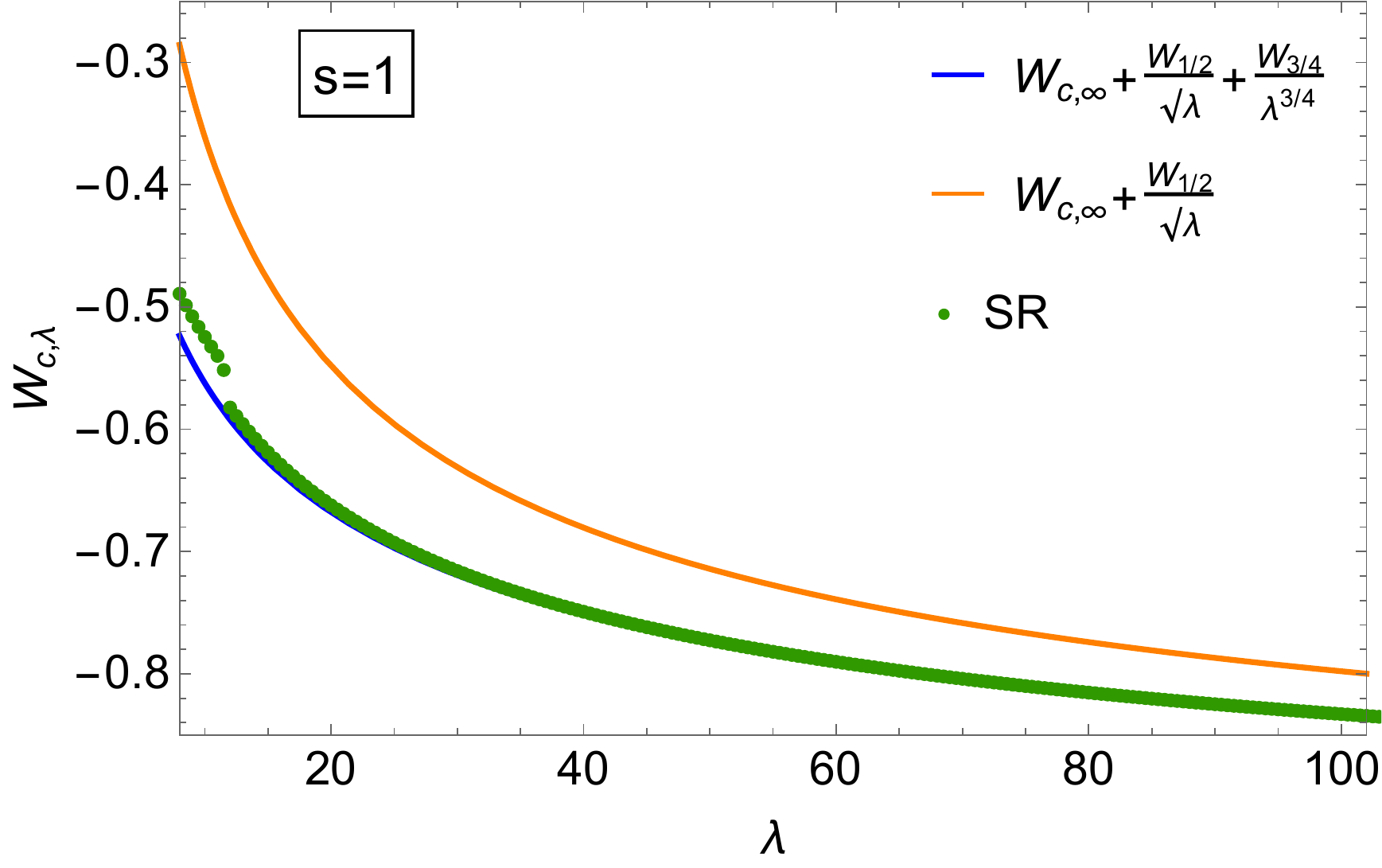}
	\includegraphics[width=0.4\textwidth]{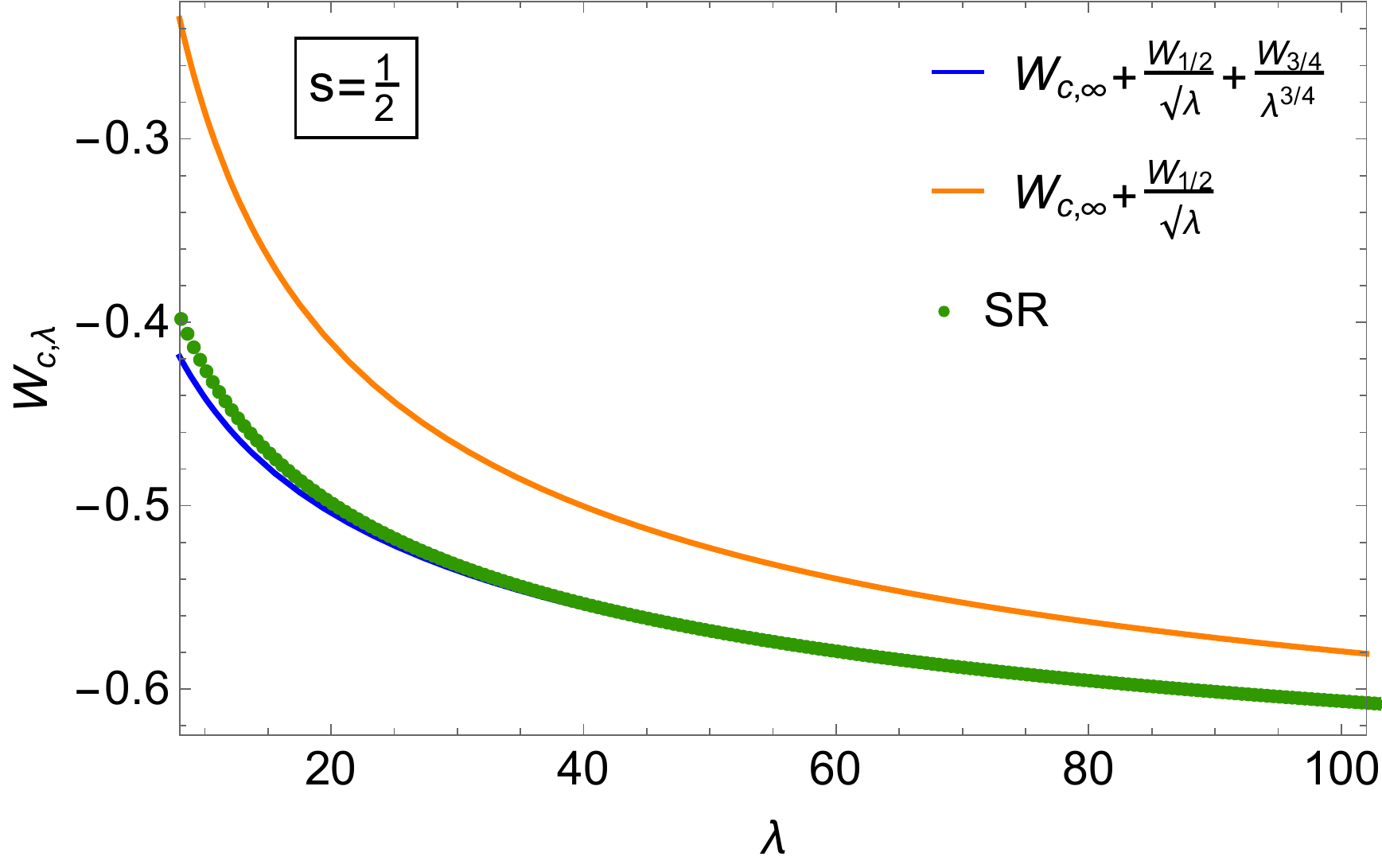}
    \caption{The large $\lambda$ expansion of Eq.~\eqref{eq:HFnonSeidl} for  H$[1,0]$ ($s=1$) and H$[\frac{1}{2},\frac{1}{2}]$ ($s=\frac{1}{2}$) compared to the spectral renormalization (SR) results for the problem \eqref{eq:SN2}.} 
    \label{fig:seriesComp}
\end{figure}

In Fig.~\ref{fig:seriesComp} we compare the expansion of Eqs.~\eqref{eq:HFnonSeidl}-\eqref{eq:lambda1/4} for $s=1$ and $s=\frac{1}{2}$ with our numerical data from the SR solution of the full $\lambda$-dependent problem \eqref{eq:SN2}, finding very good agreement for large $\lambda$.


\section{The H$_{2}$ Molecule (RHF)}\label{sec:H2}
The MP adiabatic connection for the H$_2$ molecule has been already computed by Pernal\cite{Per-JCP-18} for $0\le\lambda\le 1$ and for $R=1.4$ and $3.0$. Here we extend the calculations up to $\lambda=20$ and for other stretched geometries, up to $R=300$. The computational details are the same as in the supplementary material of Ref.~\onlinecite{VucFabGorBur-JCTC-20}, where we have used  FCI with uncontracted aug-cc-pVTZ basis set to solve the $\lambda$-dependent hamiltonian of Eq.~\eqref{eq:HlambdaHF} for the restricted HF case. 

\begin{figure}
    \centering
    \includegraphics[width=0.4\textwidth]{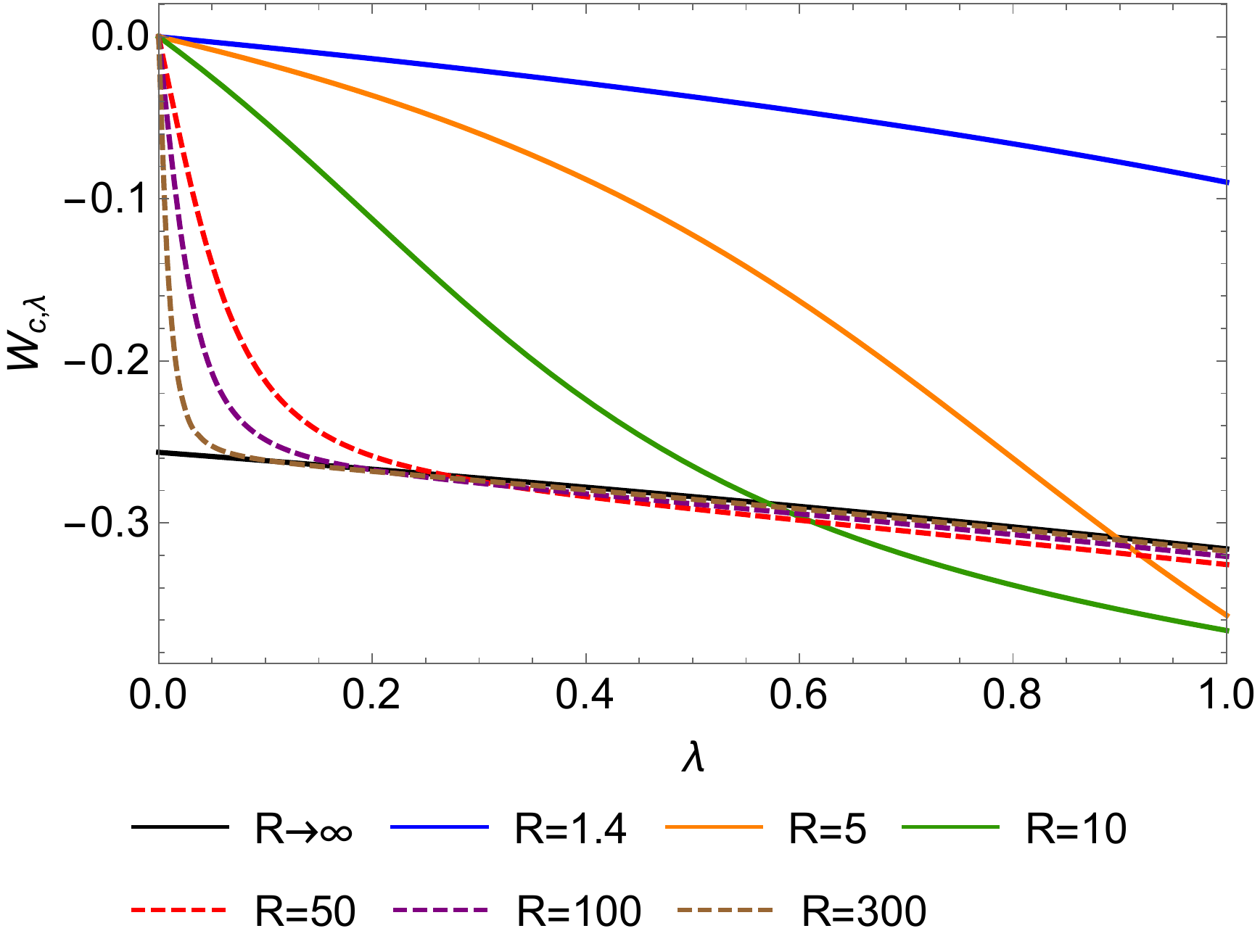}
	\includegraphics[width=0.4\textwidth]{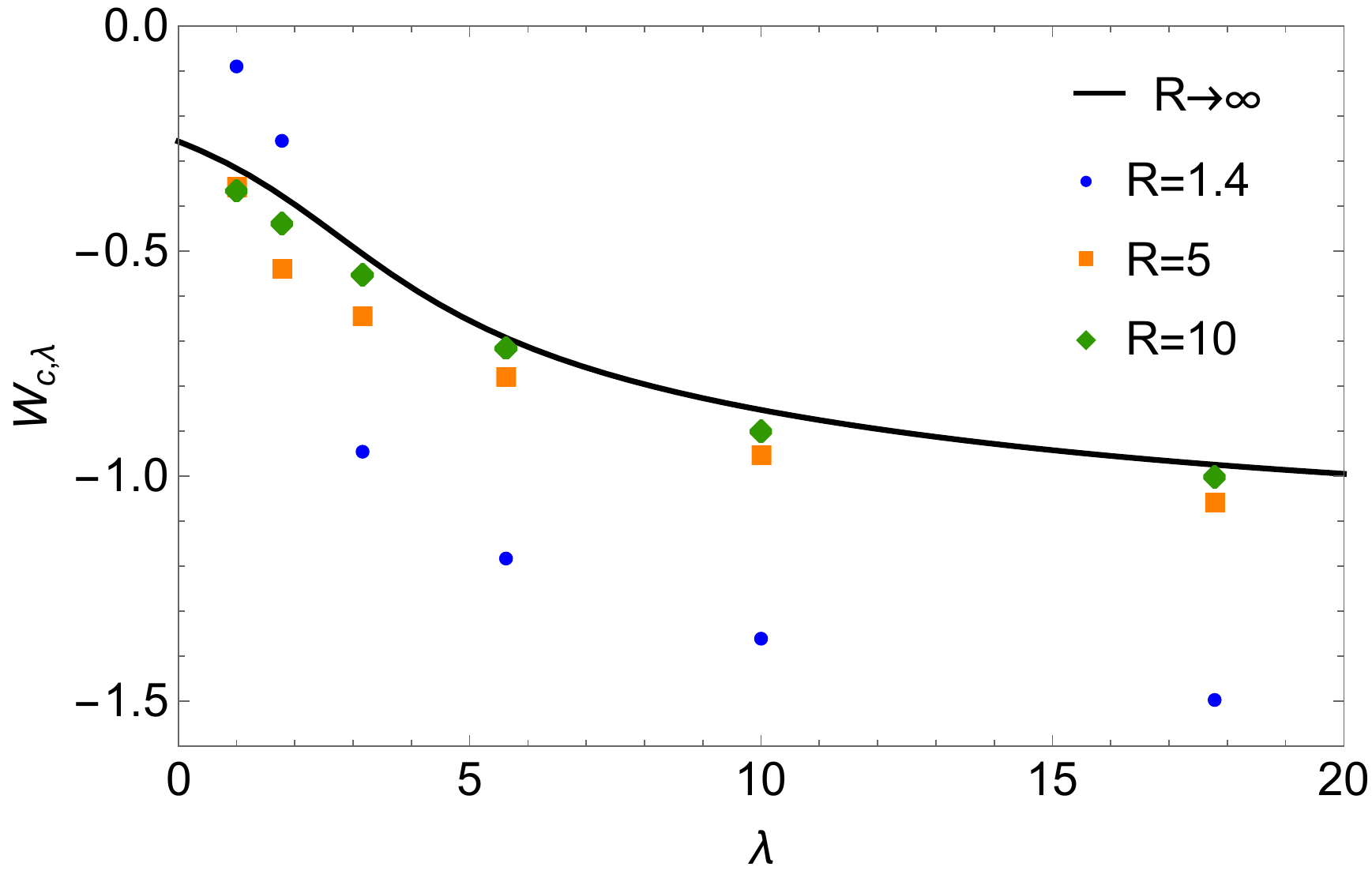}
    \caption{The $\lambda$-dependent adiabatic connection integrand  of the \ce{H2} molecule at different internuclear distances $R$ compared to twice our $W_{c,\lambda}$ for H$[\frac{1}{2},\frac{1}{2}]$ of Fig.~\ref{fig:WSTOSN12}, labeled here ``$R\rightarrow\infty$''.}
    \label{fig:H2}
\end{figure}

In Fig.~\ref{fig:H2} we compare our results for the H$_2$ molecule with twice the $W_{c,\lambda}^{\rm HF}$ for the H$[\frac{1}{2},\frac{1}{2}]$ system of Fig.~\ref{fig:WSTOSN12}. In the upper panel we first focus on $\lambda\in[0,1]$, where we see that as $R$ gets larger the H$_2$ $W_{c,\lambda}^{\rm HF}$ eventually falls on our H$[\frac{1}{2},\frac{1}{2}]$ curve, although this happens only in the extreme stretched case, with $R$ around 50 or more. 
In the lower panel we extend the $\lambda$-range up to 20, where we see that the $W_{c,\lambda}^{\rm HF}$  for $R=5$ and $R=10$ approach the H$[\frac{1}{2},\frac{1}{2}]$ curve, but at larger $\lambda$, and from below.

\begin{figure}[htp]
    \centering
    \includegraphics[width=0.4\textwidth]{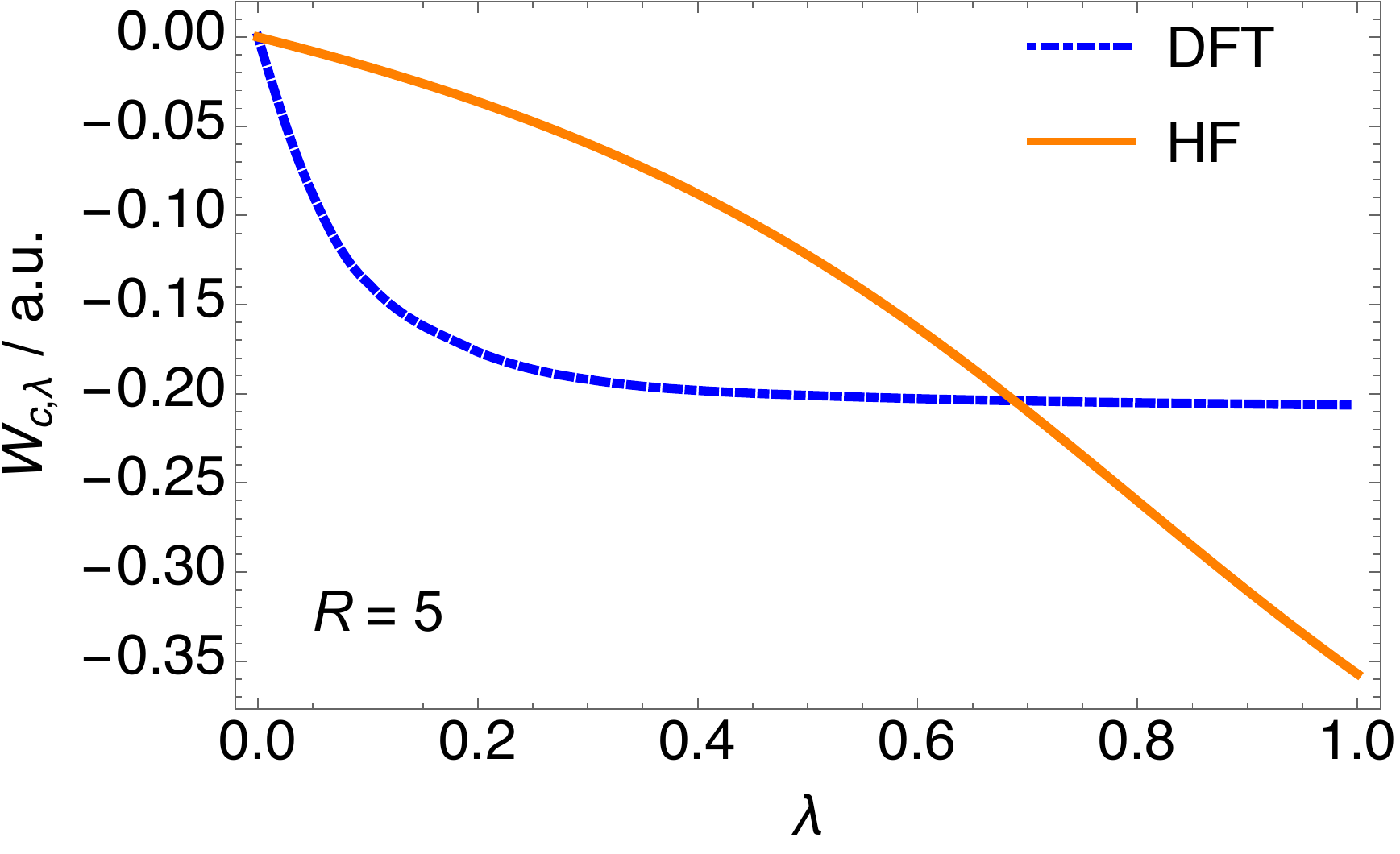}
	\includegraphics[width=0.4\textwidth]{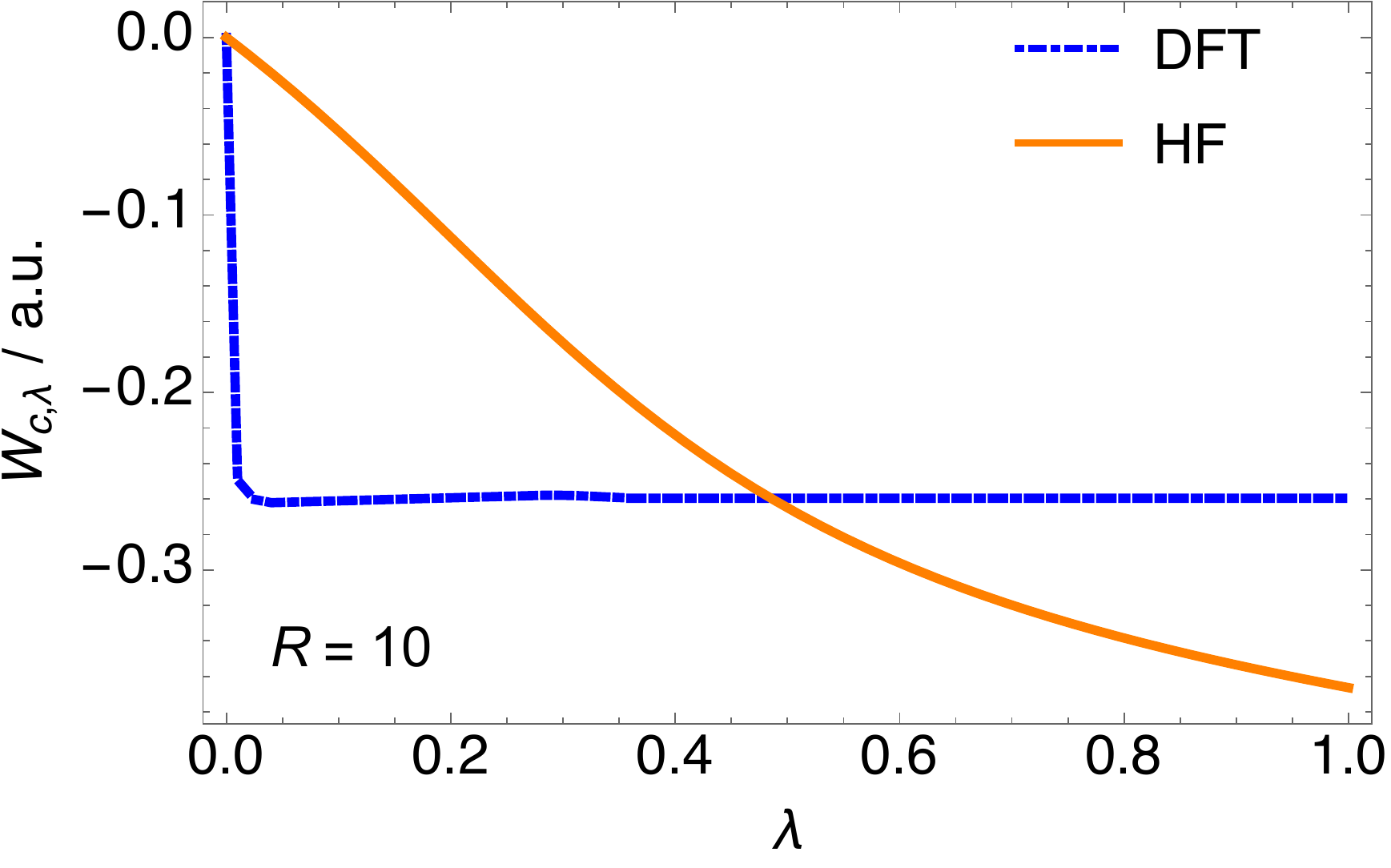}
    \caption{The $\lambda$-dependent adiabatic connection integrand for the HF  case [Eq.~\eqref{eq:HlambdaHF}] and for the density-fixed DFT case\cite{TeaCorHel-JCP-09,VucIroSavTeaGor-JCTC-16} for the \ce{H2} molecule at internuclear distance $R=5$ and $R=10$.}
    \label{fig:H2-HFDFT}
\end{figure}

This behavior is very different from the one of the density-fixed DFT adiabatic connection, in which the $\lambda\to\infty$ limit (in this case coinciding exactly with the $R\to\infty$ limit~\cite{VucWagMirGor-JCTC-15,VucIroWagTeaGor-PCCP-17}), is reached much faster as the molecule is stretched, as shown in Fig.~\ref{fig:H2-HFDFT} for the $R=5$ and $R=10$ case.

\section{From the H atom to the many-electron closed-shell case}\label{sec:HFdensfunc}
In this section we show that the result for the H atom for $s=\frac{1}{2}$ provides a variational expression for the large-$\lambda$ expansion of the MP AC in the general spin-restricted closed-shell case. 
The idea is to start from a variational ansatz more general than the one of Eq.~\eqref{eq:PsiT}, namely
\begin{equation}\label{eq:PsiT2}
	\Psi_\lambda^{h}(\rv_1,\dots,\rv_N)=\prod_{i=1}^N\loc_{i,\lambda}(|\rv_i-\rv^{\rm min}_i|),
\end{equation}
where 
\begin{equation}\label{eq:loc1}
	\loc_{i,\lambda}(r)=\lambda^{\frac{3n}{2}}\loc_i(\lambda^n\,r),
\end{equation}
$\loc_i(t)$ is a localised, normalised, 3D spherical function,
\begin{equation}
	\int d{\bf t}\, \loc_i^2(t)=1,
\end{equation}
which needs to be determined variationally.
We will set at the end $n=1/4$ in Eq.~\eqref{eq:loc1}, which is the correct scaling for the 3D case as shown in Ref.~\onlinecite{SeiGiaVucFabGor-JCP-2018}. We then evaluate the expectation of the hamiltonian $\hat{H}_{\lambda}^{\rm HF}$ of Eq.~\eqref{eq:HlambdaHF} on $\Psi_\lambda^{h}$ for a closed-shell system for large $\lambda$, where we start at $\lambda=0$ from a spin-restricted HF calculation (and for this reason, the choice of the spins is irrelevant in Eq.~\eqref{eq:PsiT2}; see also the discussion in Appendix~\ref{app:spinflipH}). The kinetic energy is simply given by
\begin{align}\label{eq:TonL}
	\langle \Psi_\lambda^{h}|\hat{T}|\Psi_\lambda^{h}\rangle=\frac{\lambda^{2n}}{2}\sum_{i=1}^N\int d{\bf t}|\nabla\loc_i(t)|^2.
\end{align}

Since when $\lambda$ is large $\Psi_\lambda^h$ localises the electrons in the minimum of the $3N$-dimensional function $\hat{V}_{ee}-\hat{J}$, we can expand it around its minimum and express it in scaled coordinates $\tv_i=\lambda^{n}(\rv_i-\rv_i^{\rm min})$,
\begin{align}
\hat{V}_{ee}-\hat{J}=C+\frac{\lambda^{-2n}}{2}\sum_{i,j=1}^N\sum_{\alpha,\beta} t_{i,\alpha}\,\mathbb{H}_{i\alpha,j\beta}\,t_{j,\beta}+O(\lambda^{-3n}),
\end{align}
where $\alpha,\beta=x,y,z$, $\mathbb{H}$ is the hessian matrix w.r.t. $\rv_i$ of the 3$N$-dimensional function $\hat{V}_{ee}-\hat{J}$ evaluated in $\rv_1^{\rm min},\dots,\rv_N^{\rm min}$, and  $C=E_{el}[\rho^{\rm HF}]-U[\rho^{\rm HF}]$ is the value of its minimum, which enters in $W_{c,\infty}^{\rm HF}$ and does not determine either $n$ or $\loc_i$.   We thus subtract $C$ and look at the term of order $\lambda^{-2n}$, whose expectation on $\Psi_\lambda^h$ gives non-zero contribution only for the diagonal terms of $\mathbb{H}$, because $\loc_i$ is spherically symmetric. Thus, we obtain, neglecting orders $\lambda^{-3n}$,
\begin{align}
\langle & \Psi_\lambda^{h}|\hat{V}_{ee}-\hat{J}|\Psi_\lambda^{h}\rangle-C=	\frac{\lambda^{-2n}}{2}\sum_{i=1}^N\sum_{\alpha}\mathbb{H}_{i\alpha,i\alpha}\int d\tv\, t_{\alpha}^2\,\loc_i^2(t) \nonumber \\
& =\lambda^{-2n}\sum_{i=1}^N 4\pi\,\rho^{\rm HF}(\rv_i^{\rm min})
\int d\tv\, \frac{t^2}{6}\,\loc_i^2(t)
, \label{eq:Vee-JonL}
\end{align}
where we have used $\int d\tv\, t_{\alpha}^2\,\loc_i^2(t)=\frac{1}{3}\int d\tv\, t^2\,\loc_i^2(t)$ and 
\begin{equation}
\sum_\alpha\mathbb{H}_{i\alpha,i\alpha}=\nabla^2_i(\hat{V}_{ee}-\hat{J})|_{\rv_i^{\rm min}}=4\pi\rho^{\rm HF}(\rv_i^{\rm min}).
\end{equation}
The expectation of $\hat{K}$ in the RHF closed-shell case is, up to orders $\lambda^{-3n}$,
\begin{widetext}
\begin{align}
	\langle \Psi_\lambda^{h}|\hat{K}|\Psi_\lambda^{h}\rangle & =\sum_{i=1}^N\int d\rv_i\int d\rv_i'\frac{\loc_{i,\lambda}(|\rv_i-\rv^{\rm min}_i|)\loc_{i,\lambda}(|\rv_i'-\rv^{\rm min}_i|)}{|\rv_i-\rv_i'|}\sum_{a=1}^{N/2}\phi_a^*(\rv_i')\phi_a(\rv_i) \nonumber \\
\label{eq:KonL}	& =\lambda^{-2n}\sum_{i=1}^N\int d\tv\int d\tv'\frac{\loc_i(t)\loc_i(t')}{|\tv-\tv'|}\underbrace{\sum_{a=1}^{N/2}|\phi_a(\rv_i^{\rm min})|^2}_{=\rho^{\rm HF}(\rv_i^{\rm min})/2}=\lambda^{-2n}\sum_{i=1}^N\frac{\rho^{\rm HF}(\rv_i^{\rm min})}{2}
	\int d\tv\int d\tv'\frac{\loc_i(t)\loc_i(t')}{|\tv-\tv'|},
\end{align}
\end{widetext}
where we have expanded the HF orbitals $\phi_a$ in scaled coordinates at large $\lambda$,
\begin{equation}
	\phi_a(\lambda^{-n}\tv_i+\rv_i^{\rm min})=\phi_a(\rv_i^{\rm min})+\lambda^{-n}\tv_i\cdot\nabla \phi_a(\rv_i^{\rm min})+O(\lambda^{-2n}).
\end{equation}
When we insert Eqs.~\eqref{eq:TonL}, \eqref{eq:Vee-JonL} and \eqref{eq:KonL} in the expectation of $\hat{H}_{\lambda}^{\rm HF}$ of Eq.~\eqref{eq:HlambdaHF} and set $n=1/4$, we obtain, neglecting orders $\lambda^{1/4}$,
\begin{align}
\langle \Psi_\lambda^{h}|\hat{H}_{\lambda}^{\rm HF}|\Psi_\lambda^{h}\rangle-\lambda C=\lambda^{1/2}\sum_{i=1}^N \tilde{E}_{\frac{1}{2}}(\rho^{\rm HF}(\rv_i^{\rm min}))[\loc_i] ,
\end{align}
where
\begin{align}
	\tilde{E}_{\frac{1}{2}}(\rho)[\loc] & =\frac{1}{2}\int d{\bf t}|\nabla\loc(t)|^2+4\pi\,\rho\,
\int d\tv\, \frac{t^2}{6}\,\loc^2(t) \nonumber \\
& +\frac{\rho}{2}
	\int d\tv\int d\tv'\frac{\loc(t)\loc(t')}{|\tv-\tv'|}.
\end{align}
Varying $\tilde{E}_{\frac{1}{2}}(\rho)[\loc]$ with respect to $\loc$ (keeping the normalisation constraint), switching to the function $ u(t)=\sqrt{4\pi}\,t\,\loc(t)$, and introducing the scaled variable $p=(4\pi\,\rho)^{1/4} t$, we obtain exactly Eq.~\eqref{eq:HFlb12} with $s=1/2$. This means that the best possible spherical variational ansatz for $\loc_i$ is the same as the one we found for the H atom, around each equilibrium position $\rv_i^{\rm min}$,
\begin{equation}
	\loc_i^{\rm opt}(t)=\frac{u_{\frac{1}{2}}(\sqrt[4]{4\pi\,\rho^{\rm HF}(\rv_i^{\rm min})}\,t)}{\sqrt{4\pi}\sqrt[4]{4\pi\,\rho^{\rm HF}(\rv_i^{\rm min})}\,t},
\end{equation}
where $u_{\frac{1}{2}}(p)/p$ is the function shown in Fig.~\ref{fig:asyWFs} for $s=1/2$. We can thus write the following general variational estimate for the functional $W_{\frac{1}{2}}[\rho^{\rm HF}]$
\begin{align}
W_{\frac{1}{2}}[\rho^{\rm HF}] & =\tilde{\epsilon}_{\frac{1}{2}}\frac{\sqrt{4\pi}}{2}\sum_{i=1}^N\left(\rho^{\rm HF}(\rv_i^{\rm min})\right)^{1/2}\nonumber \\
& =2.8687 \sum_{i=1}^N\left(\rho^{\rm HF}(\rv_i^{\rm min})\right)^{1/2}, \label{eq:W12fromH}
\end{align}
where we have used $\tilde{\epsilon}_{\frac{1}{2}}=1.6185$ from Table~\ref{tb:Channels}.

At the next leading order neither $\hat{V}_{ee}-\hat{J}$ nor $\hat{K}$ contribute because their expansion at large $\lambda$ contains only odd powers of $t_{i,\alpha}$. This is due to the fact that we only use spherical functions around $\rv_i^{\rm min}$, so we cannot probe anisotropy with our variational ansatz. The only exception is for the $\rv_i^{\rm min}$ that coincide with a nuclear position, where there is a cusp in the HF density and orbitals. In this case, exactly as in the H atom, the external potential, $\hat{V}_{ee}-\hat{J}$, and $\hat{K}$ all contribute to the same order $\lambda^{1/4}$ in the energy. We thus obtain also an estimate for the functional $W_{\frac{3}{4}}[\rho^{\rm HF}]$,
\begin{align}
W_{\frac{3}{4}}[\rho^{\rm HF}]& =\tilde{\epsilon}_{\frac{1}{4}}\frac{\sqrt[4]{4\pi}}{4} \sum_{\rv_{Z_k}}Z_k\,\left(\rho^{\rm HF}(\rv_{Z_k})\right)^{1/4} \nonumber \\
& =-1.272\sum_{\rv_{Z_k}}Z_k  \left(\rho^{\rm HF}(\rv_{Z_k})\right)^{1/4},
\end{align}
where the sum runs only over the $\rv_i^{\rm min}$ that are located at a nucleus with charge $Z_k$, and we have used the value $\tilde{\epsilon}_{\frac{1}{4}}= -2.703$ from Eq.~\eqref{eq:eps14s12}.

\section{Uniform electron gas}\label{sec:UEG}
In this section we focus on the uniform electron gas (UEG),\cite{Wig-PR-34,GiuVig-BOOK-05} which is a cornerstone in the construction of approximate density functionals, and can thus provide useful pieces of information for building models for the large-$\lambda$ limit of the MP AC. The UEG is sometimes also called jellium, although in principle the two models are defined differently:\cite{LewLieSei-JEPM-18} in the UEG there is no external potential but the electrons are constrained to have a uniform density $\rho$, while in the jellium model the external potential is fixed, determined by a background of uniform positive charge density $\rho$. However, very recently the equivalence  between the two models has been fully established, including for the strong-coupling (low-density) regime.\cite{CotPet-arxiv-17,LewLieSei-PRB-19} 
 The jellium Hamiltonian reads
\begin{align}\label{JelHam}
\hat{H}_{\mathrm{jel}}=&-\frac{1}{2}\sum_{i=1}^N\nabla^2_{\mathbf{r}_i}+\frac{1}{2}\sum_{i\neq j }^N\frac{1}{\vert\mathbf{r}_i-\mathbf{r}_j\vert}\underbrace{-\sum_{i=1}^N\int_V\frac{\rho}{\vert\mathbf{r}_i-\mathbf{r}'\vert}\mathrm{d}\mathbf{r}'}_{\hat{V}_{eb}=-\hat{J}[\rho]}\nonumber\\&+\underbrace{\frac{1}{2}\int_{V\times V}\frac{\rho^2}{\vert\mathbf{r}-\mathbf{r}'\vert}\mathrm{d}\mathbf{r}'\mathrm{d}\mathbf{r}}_{\hat{V}_{bb}=U[\rho]}
\end{align}
where we have $N$ electrons immersed in the background of positive charge density $\rho$ inside the volume $V$, and we are interested in 
the thermodynamic limit $N,V\rightarrow\infty$ with $\rho=N/V$ kept fixed, which can be done in different equivalent ways. \cite{LewLieSei-PRB-19} The relevant length scale in $\hat{H}_{\mathrm{jel}}$ is $r_s$, defined for $D=3$ as $r_s=\sqrt[3]{3/(4\pi \rho)}$: if we use scaled coordinates $\sv_i=\rv_i/r_s$, we see that the kinetic energy scales as $1/r_s^2$ while all the potential energy terms scale as $1/r_s$.
The low-density regime $r_s\rightarrow\infty$ is thus equivalent to the large-$\lambda$ case  and the electrons are believed to localize in lattice points to form a bcc Wigner crystal.\cite{Wig-PR-34,Wig-TFS-38} In the case of the UEG, the uniform density is recovered by making a linear superposition of all orientations and elementary translations of the lattice,\cite{Car-PR-61,LewLieSei-PRB-19} which is a special case of the strictly-correlated-electrons (SCE) limit\cite{SeiGorSav-PRA-07} of DFT.

The accurate large-$r_s$ treatment, carried out by Carr,\cite{Car-PR-61} leads to the expansion for the energy per electron
\begin{equation}\label{eq:WigLowExp}
\epsilon_{\mathrm{jel}}(r_s)\sim-\frac{0.896}{r_s}+\frac{1.33}{r_s^{3/2}}+O\left(\frac{1}{r_s^2}\right),\quad r_s\to\infty.
\end{equation}
The coefficient of the leading term is the Madelung constant of the bcc lattice, and it is obtained by minimising the electrostatic energy alone. The subleading term is obtained from a normal-mode analysis of zero-point oscillations of the electrons around their equilibrium positions. Notice that if instead of the normal mode calculation we use a single spherical gaussian as in the trial wave function of Eq.~\eqref{eq:PsiT}, we obtain Wigner's original result\cite{Wig-TFS-38} for the coefficient of $r_s^{-3/2}$, equal to 1.5, making an error of about 12\% with respect to the accurate 1.33. This could provide an indication of the kind of error we make when considering a spherical uncoupled approximation as we do with the trial wave function of Eq.~\eqref{eq:PsiT2}.

We thus consider the large-$\lambda$ limit of the MP adiabatic connection of Eq.~\eqref{eq:HlambdaHF}. The unrestricted HF ground-state of the UEG is never translationally invariant, even at high density, as there is always an exponentially small gain in energy with a charge- and spin-density wave.\cite{Ove-PRL-60,Ove-PRL-62,Ove-PR-68,GonHaiLew-PRA-19} Here we consider a fully restricted HF calculation, in which the translational invariance is enforced. In this simple case the electronic HF density is uniform, $\rho^{\rm HF}=\rho$ and the occupied HF orbitals are plane waves with momentum $\kv$, with  $|\kv|\le k_f$ and $k_f\coloneqq\sqrt[3]{\pi~9/4}~r_s^{-1}$. 
By comparing Eq.~\eqref{JelHam} with Eq.~\eqref{eq:EelDef} we see immediately that the leading term in Eq.~\eqref{eq:WigLowExp} is exactly equal to $E_{el}[\rho]/N$ in the thermodynamic limit,
\begin{equation}
	\lim_{\substack{N,V\to\infty \\ N/V=\rho}}\frac{E_{el}[\rho]}{N}=-\frac{0.896}{r_s},\qquad \rho=\left(\frac{4 \pi}{3}r_s^3\right)^{-1}.
\end{equation}
In other words, the unknown part $E_{el}[\rho^{\rm HF}]$ in the strong interacting limit of the restricted MP adiabatic connection for a UEG with $\dens=\dens^{\mathrm{HF}}$, is given by the leading term of the low-density expansion of the UEG. In Ref.~\onlinecite{SeiGiaVucFabGor-JCP-2018} it has been proven that for any density $\rho(\rv)$,
\begin{equation}\label{eq:ineq}
	E_{el}[\rho]\le W_\infty^{\rm DFT}[\rho],
\end{equation}
where $W_\infty^{\rm DFT}[\rho]$ is the $\lambda\to\infty$ limit of the DFT density-fixed adiabatic connection,\cite{SeiGorSav-PRA-07,GorVigSei-JCTC-09} which for the UEG corresponds to the bcc Madelung energy.\cite{GorSei-PCCP-10,LewLieSei-PRB-19} We thus see that for the case of a uniform density we have the equality 
\begin{equation}\label{eq:eq}
	E_{el}[\rho_{\rm unif}]= W_\infty^{\rm DFT}[\rho_{\rm unif}].
\end{equation}
The exact $W_\infty^{\rm DFT}[\rho]$ for a general non-uniform density $\rho(\rv)$ is involved and described by the SCE formalism.\cite{SeiGorSav-PRA-07,ButDepGor-PRA-12} It is also very well approximated by the PC model,\cite{SeiPerKur-PRA-00} which is a gradient expansion (GEA). From Eqs.~\eqref{eq:ineq}-\eqref{eq:eq} we see that in order to build a GEA for $E_{el}[\rho]$ we will most likely need a gradient correction that is {\em negative} rather than positive as it is in the PC model. This route will be pursued in future work.

By using the variational result of Eq.~\eqref{eq:W12fromH} we can obtain the adiabatic connection integrand per electron $w_{c,\lambda}^{\rm HF}$ of the  RHF hamiltonian for the UEG at large $\lambda$ as
\begin{equation}\label{eq;wcHFUEG}
	 w_{c,\lambda}^{\rm HF}=-\frac{1.354}{r_s}+\frac{1}{\sqrt{\lambda}}\frac{1.402}{r_s^{3/2}}+\dots \quad(\lambda\to\infty)
\end{equation}
where the coefficient of the $1/r_s$ term is obtained by adding $\epsilon_x=-3/4(3/2\pi)^{2/3}r_s^{-1}$ to the Madelung energy, to comply with Eq.~\eqref{eq:deriv2}. We can also see how the coefficient of the term $r_s^{-3/2}$ is raised by the operator $\hat{K}$: if we stay in a spherical approximation, without $\hat{K}$ we would obtain Wigner's result equal to $3/4=0.75$ instead of 1.402. In Appendix~\ref{app:UEG} we also report a calculation with a single gaussian including $\hat{K}$, which further illustrates the MP AC for the UEG case.

\section{Large coupling strength and strong interaction}\label{sec:SIornotSI}
In our previous work\cite{SeiGiaVucFabGor-JCP-2018} the large-$\lambda$ limit of the MP adiabatic connection defined by Eq.~\eqref{eq:HlambdaHF} was referred to as {\em strong-interaction limit}, in analogy with the density-fixed DFT adiabatic connection. However, the $\lambda\to\infty$ limit results for the H atom of Sec.~\ref{sec:Hatomresults} clash with the term strong-interaction, as this is a case in which there is no interaction at all in the exact hamiltonian. This counterintuitive result is due to the fact that in the hamiltonian \eqref{eq:HlambdaHF} it is not the full interaction operator $\hat{V}_{ee}$ that is scaled with $\lambda$, but the fluctuation potential $\hat{V}_{ee}-\hat{J}+\hat{K}$. In other words, what grows with $\lambda$ is the difference between the exact electron-electron interaction operator and its HF approximation. For a spin-polarized one-electron system this has weird consequences, which simply come from the fact that in the HF approximation the ground-state is exact, but the excited states are not. On the other hand, we have seen in Sec.~\ref{sec:UEG} that in the case of the uniform electron gas the $\lambda\to\infty$ limit of the MP AC tends to the Wigner crystal state, for which the term strong-interaction would work well. Overall, we have decided to use the term {\em large coupling strength} which seems to better describe both cases. 

\section{Conclusions and Perspectives}
We have studied the adiabatic connection of Eq.~\eqref{eq:HlambdaHF} (the MP AC) from $\lambda=0$ to $\lambda\to\infty$ for the H atom, both in the spin-polarized and spin-unpolarized case (Sec.~\ref{sec:H}). The results have revealed several interesting features of the MP AC, including an asymptotic equation for the large-$\lambda$ limit that is generally valid in a spherical approximation (as proven in Sec.~\ref{sec:HFdensfunc}). For a many-electron closed-shell system we can thus write the following large-$\lambda$ expansion
\begin{align}\label{eq:finalexp}
    W^{\rm HF}_{c,\lambda\rightarrow\infty} & = W^{\rm HF}_{c,\infty} + \frac{W^{\rm HF}_{\frac{1}{2}}}{\sqrt{\lambda}}+\frac{W^{\rm HF}_{\frac{3}{4}}}{\lambda^{\frac{3}{4}}}+\dots \\
	W^{\rm HF}_{c,\infty} & = E_{el}[\rho^{\rm HF}]+E_x \label{eq:Wcinffinal}\\
	W^{\rm HF}_{\frac{1}{2}} & \approx  2.8687 \sum_{i=1}^N\left(\rho^{\rm HF}(\rv_i^{\rm min})\right)^{1/2} \label{eq:Wc1/2fin}\\
	W^{\rm HF}_{\frac{3}{4}} & \approx -1.272\sum_{\rv_{Z_k}}Z_k  \left(\rho^{\rm HF}(\rv_{Z_k})\right)^{1/4},\label{eq:finalambda3/4}
\end{align}
where $E_{el}[\rho]$ is the electrostatic energy defined in Eq.~\eqref{eq:EelDef}, which, in turn, determines the minimising positions $\{\rv_1^{\rm min},\dots,\rv_N^{\rm min}\}$. In Eq.~\eqref{eq:finalambda3/4} the sum runs only over minimising positions that are located at a nucleus with charge $Z_k$. Equation~\eqref{eq:Wcinffinal} is exact, while Eqs.~\eqref{eq:Wc1/2fin}-\eqref{eq:finalambda3/4} are variational estimates.
We have also studied the H$_2$ molecule case (Sec.~\ref{sec:H2}) and the uniform electron gas (Sec.~\ref{sec:UEG}).

This study opens several future perspectives, for example:
\begin{itemize}
	\item The design and testing of improved interpolation formulas between the MP2 limit and the large coupling-strength limit to treat non-covalent interactions.
	\item The design of GGA's for the functionals of Eqs.~\eqref{eq:Wcinffinal}-\eqref{eq:finalambda3/4}, in a spirit similar to the PC model.\cite{SeiPerKur-PRA-00} By studying the uniform electron gas we have established here the starting point in the limit of uniform density.
	\item The generalisation of this study to other kinds of adiabatic connection appearing in wave function theory.\cite{Per-JCP-18}
\end{itemize}

\section*{Acknowledgements}
Financial support from the Netherlands Organisation for Scientific Research under Vici grant 724.017.001, the European Research Council under H2020/ERC Consolidator Grant corr-DFT (Grant Number 648932),  and H2020/MSCA-IF ``SCP-Disorder'' (Grant Number 797247) is acknowledged. SV acknowledges
funding from the Rubicon project (019.181EN.026),
which is financed by the Netherlands Organisation for Scientific
Research (NWO).

\section*{Data Availability Statement}
Data sharing not applicable. [This is a paper with only formal results, except for the numerical solution of the one-dimensional Eq.~\eqref{eq:SN2}, which can be easily reproduced, and are also available upon request to the corresponding author.]

\appendix
\section{Spin-flip in the H atom}\label{app:spinflipH}
If we allow the spin of the wave function $\Psi_\lambda$ in Sec.~\ref{sec:H} to be determined variationally, for the H$[1,0]$ case we will get a complete spin flip as soon as $\lambda>1$. In fact, the kernel of $\hat{K}$ reads explicitly (with $\xv=\rv,\sigma$, where $\sigma$ is the spin of the electron) as
\begin{equation}
	\hat{K}(\xv,\xv')=\frac{\phi_s(\rv)\phi_{s}^{*}(\textbf{r}')}{\left|\textbf{r}-\textbf{r}'\right|}|\alpha\rangle\langle\alpha'|.
\end{equation}
As long as $\lambda<1$, the lowest energy solution is attained by choosing the spin of $\Psi_\lambda$ to be $\alpha$, but 
as soon as $\lambda>1$ the expectation of $(\lambda-1)\hat{K}$ becomes positive definite, and it is thus variationally convenient to flip the spin of $\Psi_\lambda$ from $\alpha$ to $\beta$ to make it zero. In that case, the $\lambda$-dependent problem for $\lambda>1$ becomes much simpler (and less interesting) than the one treated in Sec.~\ref{sec:H}, as there is only $(1-\lambda)\,v_h(r)$ in the $\lambda$-dependent hamiltonian.

For the H$[\frac{1}{2},\frac{1}{2}]$, the spin of $\Psi_\lambda$ does not matter because the spin part of the kernel of $\hat{K}$ is simply $\frac{1}{2}(|\alpha\rangle\langle\alpha'|+|\beta\rangle\langle\beta'|)$, which has always expectation $\frac{1}{2}$. This factor is taken into account by writing $s$ in front of $\hat{K}$ in Eq.~\eqref{eq:HydrogenHF}.
The same holds for any closed-shell system in restricted HF, which is the case treated in Sec.~\ref{sec:HFdensfunc}.

\section{Explicit calculation with $\Psi_\lambda^T$ for the UEG}\label{app:UEG}
We report the explicit calculation with the simpler trial wave function of Eq.~\eqref{eq:PsiT} for the UEG HF $\lambda$-dependent hamiltonian, which we write for large $\lambda$ as
\begin{align}\label{eq:HFACjELLIUM}
\hat{H}_{\lambda\to\infty,\mathrm{jel}}^{\mathrm{HF}}=\hat{T}+\lambda\left(\hat{V}_{ee}-\hat{J}[\rho]+U[\rho]+\hat{K}\right),
\end{align}
where we have discarded terms of order $\lambda^0$, and we have included the background-background term to keep the energy per electron finite in the thermodynamic limit. The background-background will provide the term $U[\rho]$ that is anyway inside $E_{el}[\rho]$, see. Eq.~\eqref{eq:EelDef}.
Since in $D=3$ the kinetic energy and the exchange operator $\hat{K}$ enter to the same order $\sqrt{\lambda}$ in the asymptotic expansion of the energy of $\hat{\tilde{H}}_{\lambda,\mathrm{jel}}^{\mathrm{HF}}$, we expect
\begin{equation}
\frac{\min_\Psi\langle\Psi\vert\hat{\tilde{H}}_{\lambda,\mathrm{jel}}^{\mathrm{HF}}\vert\Psi\rangle}{N}+\lambda\frac{0.896}{r_s}\sim\sqrt{\lambda}~\frac{a_{\mathrm{ZP}}^{\mathrm{HF}}}{r_s^{3/2}},
\end{equation} and we aim at computing $a_{\mathrm{ZP}}^\mathrm{HF}$ variationally using the wave function of Eq.\eqref{eq:PsiT}, which we rewrite for ease of reading: 
\begin{align}\label{eq:HartreeAnsatz}
    \Psi^T_{\lambda}(\mathbf{r}_1,\dots\mathbf{r}_N)&=\prod_{i=1}^N G_{\omega_\lambda}(\mathbf{r}_i-\mathbf{r}_i^{\mathrm{min}}),\nonumber\\ G_{\omega_\lambda}(\mathbf{r})&=\bigg(\frac{\omega_{\lambda}}{\pi}\bigg)^{\frac{3}{4}}e^{-\frac{\omega_{\lambda}}{2}(\vert\mathbf{r}\vert)^2},
\end{align}
where the $\mathbf{r}_i^{\mathrm{min}}$ are the positions of the direct bcc lattice points. 

For the expectation value of the first terms of the hamiltonian \eqref{eq:HFACjELLIUM} on $\Psi_\lambda^T$ we have the standard results
\begin{align}  
	\langle \Psi_\lambda^T|\hat{T}|\Psi_\lambda^T\rangle & = N\frac{3}{4}\omega_{\lambda} \\
\langle\Psi_\lambda^T|\hat{V}_{ee}|\Psi_\lambda^T\rangle= & \frac{1}{2}\sum_{i\neq j}^N \frac{\erf\left(\sqrt{\frac{\omega_{\lambda}}{2}}|\mathbf{r}_i^{\mathrm{min}}-\mathbf{r}_j^{\mathrm{min}}|\right)}{|\mathbf{r}_i^{\mathrm{min}}-\mathbf{r}_j^{\mathrm{min}}|} \nonumber \\
& =\frac{N}{2}\sum_{\mathbf{r}_i^{\mathrm{min}}\neq {\bf 0}} \frac{\erf\left(\sqrt{\frac{\omega_{\lambda}}{2}}|\mathbf{r}_i^{\mathrm{min}}|\right)}{|\mathbf{r}_i^{\mathrm{min}}|}\\
\langle\Psi_\lambda^T| -\hat{J}[\rho]|\Psi_\lambda^T\rangle & =-\rho\sum_{i=1}^N\int_V d\rv \frac{\erf\left(\sqrt{\omega_\lambda}|\rv-\mathbf{r}_i^{\mathrm{min}}|\right)}{|\rv-\mathbf{r}_i^{\mathrm{min}}|}.
\end{align}
It is then convenient to rewrite all the $\erf$ functions as $1-\erfc$ to remove the Madelung energy, which does not depend on $\omega_\lambda$. This way, we obtain for the electrostatic part only converging integrals that lead to the original result of Wigner,\cite{Wig-TFS-38}
\begin{align}
	\frac{\langle \Psi_\lambda^T|\hat{V}_{ee}-\hat{J}[\rho]+U[\rho]|\Psi_\lambda^T\rangle}{N}+\frac{0.896}{r_s}=\rho\frac{\pi}{\omega_\lambda}+O(e^{-\omega_\lambda})
\end{align}
To evaluate the expectation of $\hat{K}$ we have to express it in terms of the HF orbitals  $\phi^{\mathrm{HF}}_j=\frac{1}{\sqrt{V}}e^{i\mathbf{k}_j\mathbf{r}}$, which yield the uniform $\rho$ via  $\rho^{\mathrm{HF}}=\rho=\sum_{j=1}^{N/2}\vert\phi^{\mathrm{HF}}_j\vert^2$, 
\begin{widetext}
\begin{align}\label{eq:KonGaus} \langle\Psi^T_{\lambda}\vert\hat{K}\vert\Psi^T_{\lambda}\rangle=&\frac{1}{V}\sum_{i=1}^N\left(\frac{\omega_{\lambda}}{\pi}\right)^{\frac{3}{2}}\sum_{\vert\mathbf{k}\vert<k_f}\int_{V}\mathrm{d}\mathbf{r}\int_{V}\mathrm{d}\mathbf{r}'\frac{e^{-\frac{\omega_{\lambda}}{2}(\vert\mathbf{r}-\mathbf{r}_i^{\mathrm{min}}\vert)^2}e^{-\frac{\omega_{\lambda}}{2}(\vert\mathbf{r}'-\mathbf{r}_i^{\mathrm{min}}\vert)^2}e^{i\mathbf{k}\cdot(\mathbf{r}-\mathbf{r}')}}{\vert\mathbf{r}-\mathbf{r}'\vert}=N\,\frac{2\left(k_f-\sqrt{\omega_{\lambda}}F_D\left(\frac{k_f}{\sqrt{\omega_{\lambda}}}\right)\right)}{\pi}
\end{align}
\end{widetext}
where $F_D(x)$ denotes the Dawson's integral.\cite{AbrSte-BOOK-72} 
This expectation value can be expanded for large $\omega_\lambda$, 
\begin{equation}
	\langle\Psi^T_{\lambda}\vert\hat{K}\vert\Psi^T_{\lambda}\rangle=N\,\rho\frac{4\pi}{\omega_\lambda}+O\left(\frac{1}{\omega_\lambda^2}\right).
\end{equation}
Putting all the terms together, the total energy per electron $\epsilon_{\lambda}^{\rm HF}= \frac{\langle\Psi_{\lambda}\vert\hat{H}_{\lambda,\mathrm{jel}}^{\mathrm{HF}}\vert\Psi_{\lambda}\rangle}{N}$ to leading orders in $\omega_\lambda$ reads 
\begin{align}\label{eq:elambdaUEGHF}
   \epsilon_{\lambda}^{\rm HF}+\lambda\frac{0.896}{r_s}  =\frac{3}{4}\omega_{\lambda}+\lambda\, \rho \frac{5\pi}{\omega_\lambda},
\end{align}
which is minimized at $\omega_\lambda=\frac{\sqrt{\lambda}}{r_s^{3/2}}\sqrt{5}\approx \sqrt{\lambda}\frac{2.24}{r_s^{3/2}}$,
yielding
\begin{equation}
    \epsilon_{\lambda}^{\rm HF}+\lambda\frac{0.896}{r_s}=\frac{\sqrt{\lambda}}{r_s^{3/2}}\frac{3}{2}\sqrt{5}\approx \sqrt{\lambda}\frac{3.354}{r_s^{3/2}}.
\end{equation}
This gives for the adiabatic connection integrand per electron a term $\lambda^{-1/2}\,r_s^{-3/2}$ with a coefficient $3.354/2=1.677$, which is, as it should, higher than the one found in Eq.~\eqref{eq;wcHFUEG} from Eq.~\eqref{eq:W12fromH}, which was equal to $1.402$.

\bibliography{bib_clean}

\end{document}